\documentclass[journal]{IEEEtran}

\normalsize

\usepackage{graphicx}
\usepackage{epstopdf}
\usepackage{amsmath,amssymb,amsfonts}

\DeclareMathOperator*{\diag}{diag}

\usepackage{algorithmic}
\usepackage{caption}
\usepackage{subcaption}
\usepackage{cite}
\usepackage{cleveref}
\usepackage{xcolor}

\begin{document}
\title{Joint CFO and Channel Estimation for RIS-aided Multi-user Massive MIMO Systems}
%
\author{Sumin~Jeong,~\IEEEmembership{Graduate Student Member,~IEEE,}
        Arman Farhang,~\IEEEmembership{Member,~IEEE,}
        Nemanja~Stefan~Perovi\'c,~\IEEEmembership{Member,~IEEE,}
        and~Mark~F.~Flanagan,~\IEEEmembership{Senior Member,~IEEE}
\thanks{S. Jeong, N. S. Perovi\'c and M. F. Flanagan are with the School of Electrical and Electronic Engineering, University College Dublin, Belfield, Dublin 4, D04 V1W8 Ireland (e-mail:
sumin.jeong@ucdconnect.ie; nemanja.stefan.perovic@ucd.ie; mark.flanagan@ieee.org).}
\thanks{A. Farhang is with the Department of Electronic and Electrical Engineering, Trinity College Dublin (TCD), College Green, Dublin 2, D02 PN40 Ireland (e-mail: arman.farhang@tcd.ie).}
\thanks{This work was supported by the Irish Research Council (IRC) under grants GOIPG/2018/2983 and IRCLA/2017/209, and by Science Foundation Ireland (SFI) under grant 19/FFP/7005. For the purpose of Open Access, the authors have applied a CC BY public copyright licence to any Author Accepted Manuscript version arising from this submission.}%
}


\maketitle
\begin{abstract}
    Accurate channel estimation is essential to achieve the performance gains promised by the use of reconfigurable intelligent surfaces (RISs) in wireless communications. In the uplink of multi-user orthogonal frequency division multiple access (OFDMA) systems, synchronization errors such as carrier frequency offsets (CFOs) can significantly degrade the channel estimation performance. This becomes more critical in RIS-aided communications, as the RIS phases are adjusted based on the channel estimates and even a small channel estimation error leads to a significant performance loss. Motivated by this, we propose a joint CFO and channel estimation method for RIS-aided multi-user massive multiple-input multiple-output (MIMO) systems. To the authors' knowledge, this represents the first work in the literature on CFO estimation for RIS-aided multi-user communication systems. Our proposed pilot structure makes it possible to accurately estimate the CFOs without multi-user interference (MUI), using the same pilot resources for both CFO estimation \emph{and} channel estimation. For joint estimation of multiple users' CFOs, a correlation-based approach is devised using the received signals at all the BS antennas. Using least-squares (LS) estimation with the obtained CFO values, the channels of all the users are jointly estimated. For optimization of the RIS phase shifts at the data transmission stage, we propose a projected gradient method (PGM) which achieves the same performance as the more computationally demanding grid search technique while requiring a significantly lower computational load. Simulation results demonstrate that the proposed method provides an improvement in the normalized mean-square error (NMSE) of channel estimation as well as in the bit error rate (BER) performance. Furthermore, we analyze the computational complexity and the pilot resource efficiency of the proposed method, and show that the proposed approach requires no extra cost in computational load or pilot overhead.
\end{abstract}

\section{Introduction}

\IEEEPARstart{M}{assive} multiple-input multiple-output (MIMO) systems utilize a large number of antennas at the base station (BS) to increase the capacity of multi-user communication networks. Compared to standard MIMO systems, massive MIMO can improve the channel capacity by orders of magnitude without requiring a larger amount of spectrum \cite{survey1_massive_MIMO, survey2_massive_MIMO, survey3_massive_MIMO}. However, the performance gains of massive MIMO are ultimately dependent on the stochastic nature of a wireless communication channel, which is in general a harsh propagation environment. Also, the blockage in mmWave bands where even massive MIMO cannot provide coverage to the users is another major issue. This blockage issue can be solved by spatially distributing many antennas \cite{survey1_RIS}. Therefore, we also consider the use of \emph{reconfigurable intelligent surfaces} (RISs) which can reduce degradation of the transmitted signal and additionally improve the system performance \cite{survey1_RIS, survey2_RIS, survey3_RIS}. The RIS is a thin metamaterial sheet which consists of a large number of passive reflecting elements. Each RIS element can control the reflections of the impinging radio waves to optimize a desirable performance metric such as the achievable rate \cite{RIS_capacity1, RIS_capacity2, RIS_capacity3}. Moreover, the RIS is a nearly-passive and highly energy-efficient structure without active electronic components (e.g. RF chains). The RIS operates in a full-duplex mode without using costly self-interference cancellation or active relaying/beamforming techniques \cite{survey1_RIS}. However, the optimal RIS reflection design requires close to perfect knowledge of the channel state information (CSI). Thus, highly accurate channel estimation is of a paramount importance in RIS-aided communication networks.  

Several channel estimation methods for RIS-assisted wireless communications have been proposed \cite{RIS_OnOff1, RIS_OnOff2, RIS_OnOff3, RIS_pattern1, RIS_pattern2, RIS_pattern3,RIS_pattern4,RIS_Active1, RIS_Active2,RIS_DL3,RIS_DL1, RIS_DL2}. In \cite{RIS_OnOff1, RIS_OnOff2, RIS_OnOff3}, on/off methods estimate the channel by switching on only one RIS element at a time. However, since multiple pilot resources are required to estimate the channel for each RIS element and the RIS usually contains a large number of reflection elements, the resulting channel estimates for all RIS elements can be outdated. On the other hand, RIS reflection pattern based methods \hspace{1sp}\cite{RIS_pattern1, RIS_pattern2, RIS_pattern3, RIS_pattern4} use a known set of well-designed RIS reflection coefficients to simultaneously estimate all channels between the BS and the RIS. In \cite{RIS_Active1, RIS_Active2, RIS_DL3}, the RIS is equipped with a small fraction of active elements which can estimate useful parameters such as angle-of-arrival (AoA) or can conduct additional signal processing at each RIS element besides passive phase shifting; however, extra hardware costs are required for using the active RIS elements. Channel estimation performance can be improved using deep learning \cite{RIS_DL3, RIS_DL1, RIS_DL2}; however, such algorithms can incur a lengthy training time. These existing channel estimation methods for RIS-assisted wireless communications have considered relevant tradeoffs between accuracy, pilot/training overhead, computational complexity, and other metrics. 

Carrier frequency offset (CFO) is an offset error between the carrier frequency of a local node and that of a reference node. If not accurately estimated and compensated, CFO can lead to significant performance degradation. This is especially true for orthogonal frequency division multiplexing (OFDM) and orthogonal frequency-division multiple access (OFDMA) systems, which are highly sensitive to the presence of CFO \cite{RIS_pattern4, OFDM_SM}. By causing inter-carrier interference (ICI) and multi-user interference (MUI), CFOs degrade the performance of channel estimation methods that assume orthogonality between different subcarriers.

In \cite{RIS_pattern4}, the authors proposed a method for CFO estimation in RIS-aided single-user OFDM systems. However, to the best of the authors' knowledge, the CFO estimation for multi-user OFDM systems equipped with an RIS has not been previously studied in the literature. 

Against this background, the contributions of this paper can be summarized as follows: 

\begin{itemize}
    \item For the first time in the literature, we investigate the effect of multiple CFOs on least-squares (LS) channel estimation methods for RIS-assisted multi-user wireless networks. 
    \item We propose a joint CFO and channel impulse response (CIR) estimation method that does not require any additional signaling overhead for CFO estimation than the pilot sequences that are utilized for channel estimation.
    \item To evaluate the performance of the proposed estimation method, we compare it to the CIR estimation method for OFDMA systems in \cite{RIS_pattern3} and the joint CFO and CIR estimation method in \cite{RIS_pattern4} with time-division multiple access (TDMA).
    \item We propose a projected gradient method (PGM) for optimizing the RIS phase shifts after CFO estimation. The proposed PGM requires a significantly lower computational complexity to optimize the RIS phase shifts.
\end{itemize}

\textit{Notation}: Lowercase bold symbols denote column vectors; uppercase bold symbols denote matrices. Superscripts $(\cdot)^{\rm{T}}$, $(\cdot)^{\rm{H}}$ and $(\cdot)^{-1}$ denote matrix transpose, Hermitian transpose, and inversion operations, respectively. $\boldsymbol{I}_a$, $\boldsymbol{0}_{1\times q}$ and $\boldsymbol{0}_{p\times q}$ denote an $a \times a$ identity matrix, an $1\times q$ zero vector and a $p\times q$ zero matrix, respectively. $\diag \{\boldsymbol{x}\}$ denotes a diagonal matrix with diagonal entries equal to those of vector $\boldsymbol{x}$. $\Vert \boldsymbol{A} \Vert$ denotes the Frobenius norm of the matrix $\boldsymbol{A}$. $((a))_{b}$ denotes an operation of $a$ modulo $b$. $\mathbb{E}\{X\}$, Var$(\cdot)$, $\nabla$, and $\angle$ denote expectation, variance, gradient, and angle operators, respectively.

\section{System model}

\begin{figure}[t]
    \centering
    \includegraphics[scale=0.15]{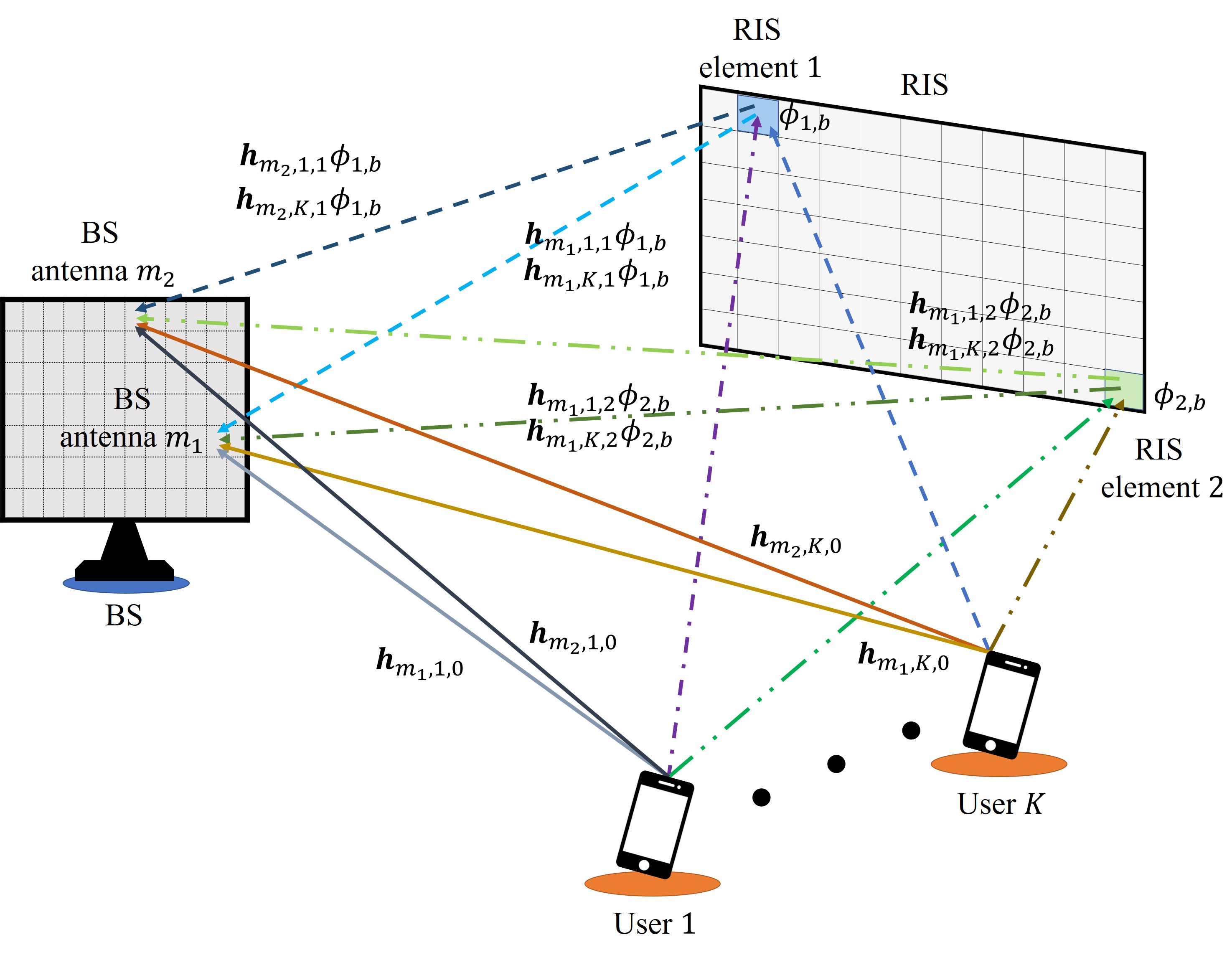}
    \caption{RIS-aided system comprised of a multiple-antenna BS and $K$ single-antenna users.}
    \label{Network_diagram}
\end{figure}

In this paper, we consider an RIS-assisted OFDM system with $N$ subcarriers transmitting over frequency-selective fading channels. As shown in Fig. \ref{Network_diagram}, the RIS is deployed to enable uplink (UL) communication from $K$ single-antenna users to a BS equipped with $M$ antennas. The RIS consists of $R$ passive reflecting elements, each of which can independently adjust the phase of the reflected signal. BS antennas are closely collocated and share the same frequency oscillator. Therefore, the transmission between the given user $k$ and any BS antenna exhibits the same CFO; we denote this CFO by $\epsilon_{k}$ (note that this is independent of the BS antenna index $m$).

Between each user and each BS antenna, there is one direct path from the user to the BS antenna and there are also $R$ reflected paths via the RIS elements. Therefore, the total number of channel paths between any user and any BS antenna is $R+1$. We assume that all channels exhibit frequency-selective fading and that the baseband equivalent channels between the users and the BS have a delay spread of at most $L$ samples. $\boldsymbol{g}_{k,m,r} = [g_{k,m,r} (0), g_{k,m,r} (1), \ldots, g_{k,m,r} (L-1)]^{\rm T}$ represents the CIR from user $k$ to BS antenna $m$ via the $r$-th element of the RIS. $\boldsymbol{h}_{k,m,r} =[h_{k,m,r} (0), h_{k,m,r} (1), \ldots, h_{k,m,r} (N-1)]^{\rm{T}} = \boldsymbol{F}_{N,L} \boldsymbol{g}_{k,m,r}$ represents the corresponding channel frequency response (CFR), where $\boldsymbol{F}_N=[\boldsymbol{f}_{0},\boldsymbol{f}_{1},\ldots,\boldsymbol{f}_{N-1}]$ denotes the $N \times N$ unitary discrete Fourier transform (DFT) matrix, whose elements are given as  $[\boldsymbol{F}_N]_{p,q} = \frac{1}{\sqrt{N}} e^{-j\frac{2\pi p q}{N}}$ for $p,q \in \{ 0, 1, \ldots, N-1 \}$ and $\boldsymbol{F}_{N,L}=[\boldsymbol{f}_{0}, \boldsymbol{f}_{1}, \ldots, \boldsymbol{f}_{L-1}]$.

In order to estimate the $KM(R+1)$ channel response vectors, the pilot sequence frame is divided into $R+1$ blocks. In pilot block $b \in \{ 0,1,\ldots, R \}$, the RIS reflection coefficient vector $\boldsymbol{\phi}_b=[\phi_{0,b}, \phi_{1,b}, \ldots, \phi_{R,b}]^{\rm T}$ is assigned. Path $0$ represents the direct path which corresponds to an RIS reflection coefficient of unity, i.e., $\phi_{0,b}=1$ $\forall$ $b$. In this paper, we assume ideal (i.e., lossless) signal reflection, which means $\vert \phi_{r,b} \vert = 1$ $\forall$ $r$ and $b$. The corresponding RIS reflection coefficient matrix is denoted by $\boldsymbol{\Phi} = [\boldsymbol{\phi}_0, \boldsymbol{\phi}_1, \ldots, \boldsymbol{\phi}_{R}]$. In order to use a LS channel estimation method, we assume $\boldsymbol{\Phi}^{\rm{H}}\boldsymbol{\Phi}=\boldsymbol{I}_{R+1}$.

In pilot block $b \in \{ 0,1,\ldots, R \}$, user $k$ transmits the time-domain pilot sequence $\boldsymbol{x}_{k,b} = [x_{k,b}(0), x_{k,b}(1), \ldots, x_{k,b}(N-1)]^{\rm T}$. This vector is the OFDM modulated signal, where OFDM modulation can be represented as the multiplication of the IDFT matrix by the frequency-domain pilot vector. The corresponding frequency-domain pilot signal is defined as $\boldsymbol{s}_{k,b} = \boldsymbol{F}_N \boldsymbol{x}_{k,b} = [s_{k,b}(0), s_{k,b}(1), \ldots, s_{k,b}(N-1)]^{\rm T}$. At the transmitter of user $k$, a cyclic prefix (CP) of length $L_{\rm{CP}}$ is added to the time-domain sequence, i.e., $x_{k,b}(u)=x_{k,b}( u+L_{\rm{CP}} )$ for $-L_{\rm{CP}} \le u \le -1$. We consider the CP to be long enough to accommodate small timing offsets between different user signals, i.e., the users are quasi-synchronous in time. The resulting time-domain signal is transmitted to the BS. After the CP removal, sample $u$ of the time-domain received signal at BS antenna $m$ in pilot block $b$ can be written as \cite{OFDM_SM}
\begin{equation}\label{eq:y_k_n}
\begin{split}
    y_{m,b}(u)=&
        \sum_{k=1}^K
        \sum_{r=0}^{R}
        \sum_{l=0}^{L-1}
            e^{j\frac{2\pi\epsilon_k (b L_{\rm{s}}+u)}{N}}
            x_{k,b} (((u-l))_{N})
            g_{k,m,r}(l)    
            \\
            &\times
            \phi_{r,b}
        +v_{m,b} (u)
        \\
        =&
        \sum_{k=1}^K
            e^{j\frac{2\pi \epsilon_k (b L_{\rm{s}}+u)}{N}}
            \widetilde{y}_{k,m,b}(u)
        +v_{m,b}(u)
        ,
\end{split}
\end{equation}
where $\epsilon_{k} \in (-0.5,0.5]$ denotes the CFO for user $k$ normalized by the subcarrier spacing, $L_{\rm{s}}=L_{\rm{CP}}+N$, $\widetilde{y}_{k,m,b}(u) = \sum_{l=0}^{L-1} x_{k,b}( ((u-l))_{N} ) \bar{g}_{k,m,b}(l)$, $\bar{g}_{k,m,b}(l)=\sum_{r=0}^{R} g_{k,m,r}(l) \phi_{r,b}$, and  $v_{m,b}(u)$ is time-domain circularly-symmetric complex additive white Gaussian noise (AWGN) having zero mean and variance $\sigma^2$. Also, we define vectors $\boldsymbol{y}_{m,b}=[y_{m,b}(0), y_{m,b}(1), \ldots, y_{m,b}(N-1)]^{\rm{T}}$ for each $m$ and $b$.

After discarding the CP and performing $N$-point DFT operations on the received signal at each BS antenna, the corresponding frequency-domain received signal at BS antenna $m$ on subcarrier $n$ for pilot block $b$ can be written as \cite{OFDM_SM}
\begin{equation}\label{eq:r_k_n}
\begin{split}
    r_{m,b}(n)=&
    \sum_{k=1}^K
    \sum_{r=0}^{R}
    \sum_{p=0}^{N-1}
        e^{j\frac{2\pi \epsilon_{k} b L_{\rm{s}}}{N}}
        s_{k,b}(p)      f_{\text{s}}(p-n+\epsilon_{k})
        \\
        &\times
        h_{k,m,r}(p)    
        \phi_{r,b}
    +w_{m,b}(n)
    ,
\end{split}
\end{equation}
where $f_{\rm{s}}(a)=\frac{\sin{(\pi a)}}{N\sin{(\pi a/N)}} e^{j\pi\frac{N-1}{N} a}$ represents the CFO effect in the frequency domain, $\boldsymbol{w}_{m,b} = \boldsymbol{F}_N \boldsymbol{v}_{m,b} = [w_{m,b}(0), w_{m,b}(1), \ldots, w_{m,b}(N-1)]^{\rm{T}}$, and $\boldsymbol{v}_{m,b} = [v_{m,b}(0), v_{m,b}(1), \ldots, v_{m,b}(N-1)]^{\rm{T}}$. Consequently, the frequency-domain received vector at BS antenna $m$ for pilot block $b$ can be written as $\boldsymbol{r}_{m,b}=[r_{m,b}(0), r_{m,b}(1), \ldots, r_{m,b}(N-1)]^{\rm{T}}$.

\section{CFO Effect on Channel Estimation}

In this section, we provide a brief overview of two LS channel estimation methods for RIS-assisted systems in the multi-user scenario. We demonstrate that for these methods, even small CFO values can significantly affect the channel estimation performance.

\subsection{CIR estimation method using OFDMA \cite{RIS_pattern3}}

A frequency-domain LS channel estimation method for an RIS-aided multi-user OFDMA system without any CFO was proposed in \cite{RIS_pattern3}. In order to avoid MUI, disjoint pilot tone allocations are utilized for all users, i.e., each subcarrier at each block is allocated to only one user. The same number of subcarriers $N_{\rm{s}}=N/K\in\mathbb{Z}$ is assigned to each user. Also, the authors of \cite{RIS_pattern3} assume $N_{\rm{s}}=L$ as the minimum required number of subcarriers assigned to one user. The authors consider equal transmit power allocation for each user over the assigned subcarrier subset for pilot block $b$. Moreover, an equal power is allocated to each subcarrier. 

If $\epsilon_k=0$ $\forall$ $k$, the frequency-domain received signal can be rewritten as
\begin{equation}\label{eq:r_mb_n_no_CFO}
    r_{m,b}(n)=
    \sum_{k=1}^K
    \sum_{r=0}^{R}
        s_{k,b}(n)
        h_{k,m,r}(n)    \phi_{r,b}
    +w_{m,b}(n)
    .
\end{equation}
For $N_{\rm{s}}=L$ and interleaved pilot allocation as in \cite{RIS_pattern3}, the frequency-domain pilot symbol vector from user $k$ for pilot block $b$ after subcarrier mapping can be rewritten as
$\boldsymbol{s}_{k,b} =\boldsymbol{\Gamma}_{k,b} \tilde{\boldsymbol{s}}_{k,b}$, where $\tilde{\boldsymbol{s}}_{k,b}=[\tilde{s}_{k,b}(0), \tilde{s}_{k,b}(1), \ldots, \tilde{s}_{k,b}(L-1)]^{\rm{T}}$ represents the frequency-domain pilot sequence of length $L$ from user $k$ for pilot block $b$ and $\boldsymbol{\Gamma}_{k,b}$ is the $N\times L$ subcarrier allocation matrix of user $k$ for pilot block $b$. $\boldsymbol{\Gamma}_{k,b}$ is comprised of the columns of the identity matrix whose indices belong to the subcarrier set assigned to user $k$ \cite{RIS_pattern3}.

Because of the disjoint pilot tone allocations, the received signal vectors for user $k$ at BS antenna $m$ can be written as
\begin{equation}\label{eq:r_kmb_no_CFO}
    \boldsymbol{r}_{k,m,b}=
    \boldsymbol{\Gamma}_{k,b}^{\rm{T}}
    \boldsymbol{r}_{m,b}
    .
\end{equation}

When the same disjoint pilot tones and pilot sequences are assigned over all pilot blocks, i.e., $\boldsymbol{\Gamma}_{k,b} = \boldsymbol{\Gamma}_{k}$ and $\boldsymbol{s}_{k,b} = \boldsymbol{s}_{k}$ for all $b$, the frequency-domain received signal matrix from user $k$ at BS antenna $m$ can be written as
\begin{equation}\label{eq:R_km_no_CFO}
\begin{split}
    \boldsymbol{R}_{k,m}
    =&
    \boldsymbol{\Gamma}_{k}^{\rm{T}}
    \diag\{ \boldsymbol{s}_{k} \}
    \boldsymbol{F}_{N,L}
    \boldsymbol{G}_{k,m}
    \boldsymbol{\Phi}
    +
    \boldsymbol{W}_{k,m}
    \\
    =&
    \boldsymbol{\Lambda}_{k}
    \boldsymbol{G}_{k,m}
    \boldsymbol{\Phi}
    +
    \boldsymbol{W}_{k,m}
    ,
\end{split}
\end{equation}
where $\boldsymbol{R}_{k,m}=[\boldsymbol{r}_{k,m,0}, \boldsymbol{r}_{k,m,1}, \ldots, \boldsymbol{r}_{k,m,R}]$, $\boldsymbol{\Lambda}_{k} = \boldsymbol{\Gamma}_{k}^{\rm{T}} \diag\{ \boldsymbol{s}_{k} \} \boldsymbol{F}_{N,L}$, $\boldsymbol{G}_{k,m}=[\boldsymbol{g}_{k,m,0}, \boldsymbol{g}_{k,m,1}, \ldots,$ $\boldsymbol{g}_{k,m,R}]$, $\boldsymbol{W}_{k,m}=\boldsymbol{\Gamma}_{k,b}^{\rm{T}} \boldsymbol{W}_{m}$, $\boldsymbol{W}_{m}=[\boldsymbol{w}_{m,0}, \boldsymbol{w}_{m,1}, \ldots, \boldsymbol{w}_{m,R}]$, and $\boldsymbol{w}_{m,r}=[w_{m,r}(0), w_{m,r}(1), \ldots,$ $ w_{m,r}(N-1)]^{\rm{T}}$.

When rank$(\boldsymbol{\Lambda}_{k})=L$, i.e., $N_{\rm{s}} \ge L$, and $\boldsymbol{\Phi}$ is an $(R+1)\times (R+1)$ DFT matrix, the CIR matrix between user $k$ and BS antenna $m$ can be estimated as
\begin{equation}\label{eq:hat_G_km_no_CFO}
    \widehat{\boldsymbol{G}}_{k,m}=
    \boldsymbol{\Lambda}_{k}^{\dagger}
    \boldsymbol{R}_{k,m}
    \boldsymbol{\Phi}^{-1}
    = 
    \boldsymbol{G}_{k,m}
    +
    \boldsymbol{\Lambda}_{k}^{\dagger}
    \boldsymbol{W}_{k,m}
    \boldsymbol{\Phi}^{-1}
    ,
\end{equation}
where $\boldsymbol{\Lambda}_{k}^{\dagger}=(\boldsymbol{\Lambda}_{k}^{\rm{H}} \boldsymbol{\Lambda}_{k})^{-1} \boldsymbol{\Lambda}_{k}^{\rm{H}}$ denotes the left pseudo-inverse of $\boldsymbol{\Lambda}_{k}$.

As described above, by using OFDMA, the CIR matrix for each user can be individually estimated. However, this method is designed without considering CFOs. Since the CFOs break the orthogonality of the subcarriers, there will be MUI from other users when we estimate $\boldsymbol{G}_{k,m}$ for user $k$. In the presence of multiple CFOs, expanding (\ref{eq:r_kmb_no_CFO}), the received signal vectors for user $k$ at BS antenna $m$ can be obtained as
\begin{equation}\label{eq:r_kmb_CFO}
\begin{split}
    \boldsymbol{r}_{k,m,b}=
    \boldsymbol{\Gamma}_{k}^{\rm{T}}
    \sum_{q=1}^K
    e^{j\frac{2\pi \epsilon_{q} b L_{\rm{s}}}{N}}
    \boldsymbol{\Pi}_{q}
    \boldsymbol{\Gamma}_{q}
    \diag\{ \tilde{\boldsymbol{s}}_{q} \}
    \boldsymbol{\bar{h}}_{q,m,b}
    +
    \boldsymbol{\widetilde{w}}_{k,m,b}
    ,
\end{split}
\end{equation}
where $\boldsymbol{\bar{h}}_{k,m,b}=[\bar{h}_{k,m,b}(0), \bar{h}_{k,m,b}(1), \ldots, \bar{h}_{k,m,b}(N-1)]^{\rm{T}}$, $\bar{h}_{k,m,b}(n)=\sum_{r=0}^{R} h_{k,m,r}(n) \phi_{r,b}$, $\boldsymbol{\widetilde{w}}_{k,m,b}$ $ =\boldsymbol{\Gamma}_{k}\boldsymbol{w}_{m,b}$, $\boldsymbol{w}_{m,b}=[w_{m,b}(0), w_{m,b}(1), \ldots, w_{m,b}(N-1)]^{\rm{T}}$, and $\boldsymbol{\Pi}_{q}$ is a circulant matrix with first row equal to $[f_{\text{s}}(\epsilon_{q}), f_{\text{s}}(\epsilon_{q}-1), \ldots, f_{\text{s}}(\epsilon_{q}-N+1)]^{\rm T}$.

Hence, the CIR estimate vector for pilot block $b$ in the presence of CFO can be
\begin{equation}
\begin{split}
    \widehat{\boldsymbol{\tilde{g}}}_{k,m,b}=& \;
    \boldsymbol{\Lambda}_{k}^{\dagger}
    \boldsymbol{r}_{k,m,b}
    =
    (\boldsymbol{I}_{L}+\boldsymbol{\Delta}_{\Lambda,k})
    e^{j\frac{2\pi \epsilon_{k} b L_{\rm{s}}}{N}}
    \boldsymbol{\bar{g}}_{k,m,b}
    +
    \boldsymbol{\delta}_{\rm{MUI},k}
    \\&
    +
    \boldsymbol{\Lambda}_{k}^{\dagger}
    \boldsymbol{\widetilde{w}}_{k,m,b}
    \\
    =&
    e^{j\frac{2\pi \epsilon_{k} b L_{\rm{s}}}{N}}
    \boldsymbol{\bar{g}}_{k,m,b}
    +
    \boldsymbol{\varpi}_{k,m,b}
    ,
\end{split}
\end{equation}
where $\boldsymbol{\bar{g}}_{k,m,b} = [ \bar{g}_{k,m,b}(0), \bar{g}_{k,m,b}(1), \ldots, \bar{g}_{k,m,b}(L-1)]^{\rm{T}}$ and $\boldsymbol{\varpi}_{k}=\boldsymbol{\delta}_{\Lambda,k} e^{j\frac{2\pi \epsilon_{k} b L_{\rm{s}}}{N}} \boldsymbol{\bar{g}}_{k,m,b} + \boldsymbol{\delta}_{\rm{MUI},k}+\boldsymbol{\Lambda}_{k}^{\dagger}\boldsymbol{\widetilde{w}}_{k,m,b}$ is the total error vector of length $L$ which includes both the interference term and the noise term for user $k$. Here, $\boldsymbol{\Delta}_{\Lambda,k}=\boldsymbol{\Lambda}_{k}^{\dagger} \boldsymbol{\Pi}_{k} \boldsymbol{\Gamma}_{k} \diag\{ \boldsymbol{s}_{k} \} \boldsymbol{F}_{N,L}-\boldsymbol{I}_{L}$ is the error matrix of size $L\times L$, which causes ICI for user $k$ (due to $\boldsymbol{\Pi}_{k}$). $\boldsymbol{\delta}_{\rm{MUI},k}$ is the $L\times 1$ vector of the MUI  for user $k$ from other users (due to $\boldsymbol{\Pi}_{q}$, where $q\neq k$).

Consequently, the CIR matrix for user $k$ at BS antenna $m$ is estimated as
\begin{equation}\label{CIR_est_ref}
\begin{split}
    \widehat{\boldsymbol{G}}_{k,m}=& \;
    \boldsymbol{\widehat{\tilde{G}}}_{k,m}
    \boldsymbol{\Phi}^{-1}
    =
    \boldsymbol{G}_{k,m}
    \boldsymbol{\widetilde{\Phi}}
    \boldsymbol{\Phi}^{-1}
    +
    \boldsymbol{\Omega}_{k,m}
    \boldsymbol{\Phi}^{-1}
    ,
\end{split}
\end{equation}
where $\boldsymbol{\widehat{\tilde{G}}}_{k,m}= [ \boldsymbol{\widehat{\tilde{g}}}_{k,m,0}, \boldsymbol{\widehat{\tilde{g}}}_{k,m,1}, \ldots, \boldsymbol{\widehat{\tilde{g}}}_{k,m,R} ]$, 
$\boldsymbol{\widetilde{\Phi}}=[ \boldsymbol{\phi}_{0}, e^{j\frac{2\pi \epsilon_{k} L_{\rm{s}}}{N}}\boldsymbol{\phi}_{1}, \ldots, e^{j\frac{2\pi \epsilon_{k} R L_{\rm{s}}}{N}}\boldsymbol{\phi}_{R} ]$, 
and $\boldsymbol{\Omega}_{k,m}= [ \boldsymbol{\varpi}_{k,m,0}, \boldsymbol{\varpi}_{k,m,1}, \ldots, \boldsymbol{\varpi}_{k,m,R} ]$.

\begin{figure}[t]
    \centering
    \includegraphics[scale=0.5]{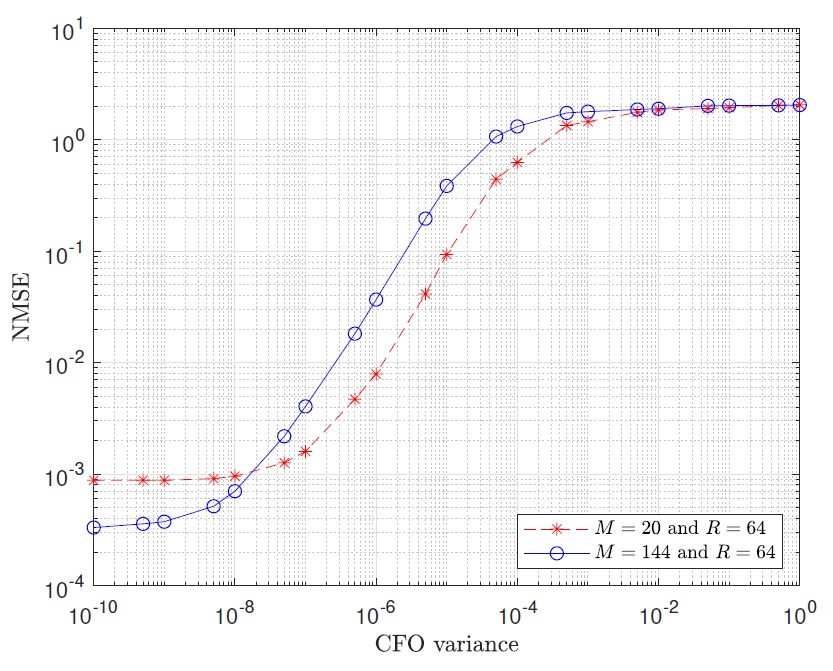}
    \caption{NMSE performance of the CIR estimation in \cite{RIS_pattern3} as a function of the CFO variance, for different values of $M$ and $R$. Here the SNR is 10 dB, $L = 32$, and $K = 3$.}
    \label{NMSE_CIR_as_CFO}
\end{figure}

In Fig. \ref{NMSE_CIR_as_CFO}, the NMSE performance of the CIR estimation method using OFDMA is shown as a function of the variance of the CFOs (this variance is assumed to be the same for each CFO). Here, we define NMSE as NMSE$=\mathbb{E} \Vert \boldsymbol{G}_{k,m} - \hat{\boldsymbol{G}}_{k,m} \Vert^2 / K \Vert \boldsymbol{G}_{k,m} \Vert^2$. In this numerical result, the interleaved subcarrier allocation is utilized. The NMSE performance becomes worse with the increase of the CFO range, and even a small value of the CFO variance can significantly degrade the accuracy of the channel estimation method in \cite{RIS_pattern3}. Note that a CFO variance of $10^{-4}$ corresponds to CFOs in the range of $10^{-2}$, which has been shown in the OFDM literature to have little effect on the performance \cite{OFDM_CFO}. However, for RIS-aided OFDM systems, it has a highly detrimental effect on the CIR estimation accuracy. The reasons for this behavior are as follows: 1) when $\epsilon_{k} \neq 0$, the term $e^{\frac{j 2 \pi \epsilon_{q} b L_{\rm{S}}}{N}}$ in $\boldsymbol{\widetilde{\Phi}}$ disables accurate estimation of $\boldsymbol{G}_{k,m}$, since $\boldsymbol{\widetilde{\Phi}} \boldsymbol{\Phi}^{-1} \neq \boldsymbol{I}_{R+1}$; and 2) ICI and MUI, besides the noise, are absorbed into $\boldsymbol{\Omega}_{k,m} \boldsymbol{\Phi}^{-1}$, so the error terms in (\ref{CIR_est_ref}) degrade the accuracy more substantially than the noise term in (\ref{eq:hat_G_km_no_CFO}). Consequently, the entire set of all CFOs needs to be estimated and compensated.

\subsection{Joint CFO and CIR estimation method using TDMA \cite{RIS_pattern4}}

\begin{figure}[t]
    \centering
    \includegraphics[scale=0.35]{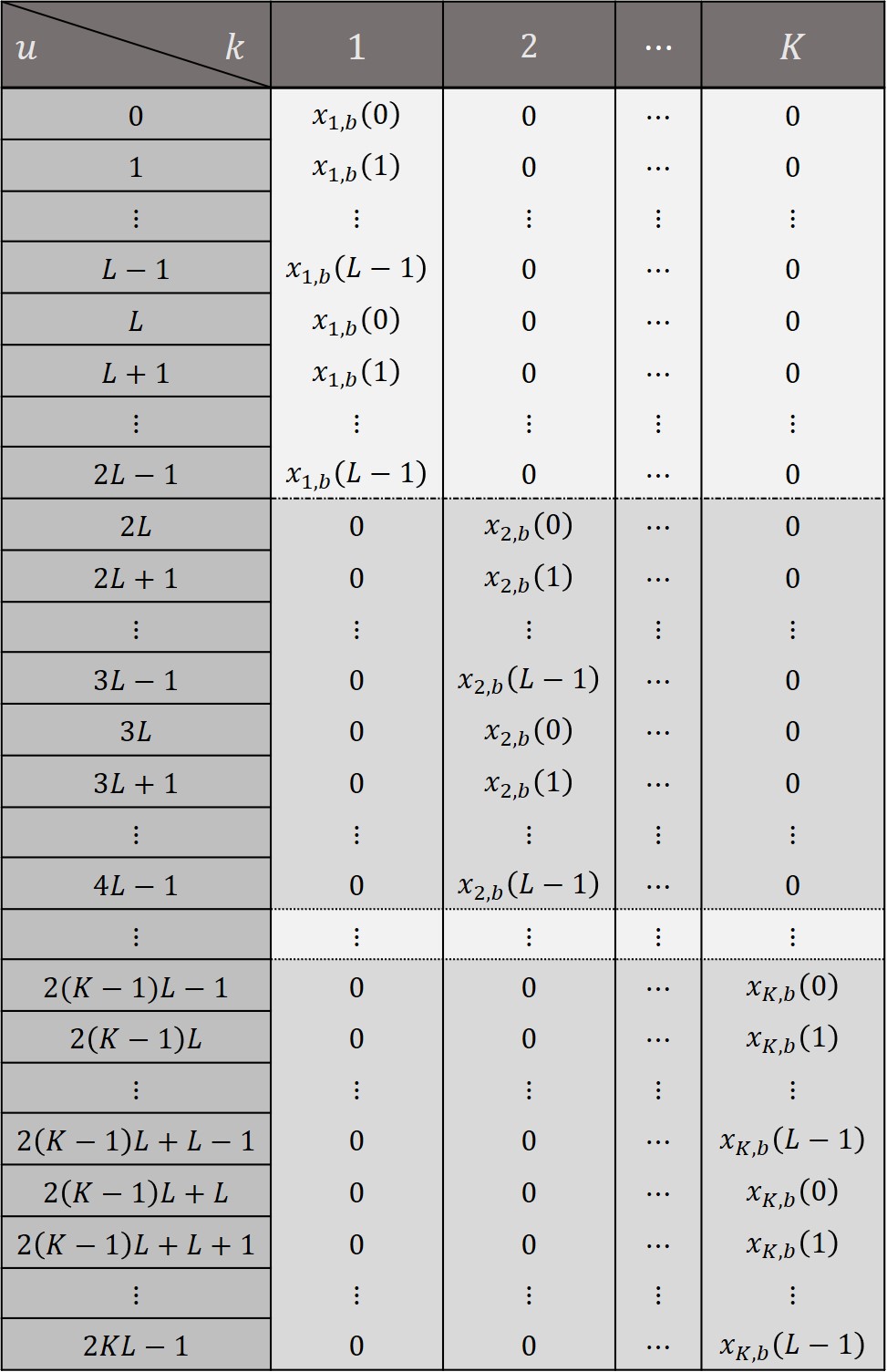}
    \caption{TDMA pilot structure in the time domain for pilot block $b$ when $N=2KL$.}
    \label{TDMA_PS_user_k}
\end{figure}

As shown in the previous subsection, the MUI caused by CFOs can seriously degrade the performance of CIR estimation in OFDMA systems. This is because the frequency-domain orthogonality between subcarrier subsets is highly sensitive to CFOs. To tackle this issue, TDMA pilot sequences can be employed and CFO/channel estimation performed in the time domain (see Fig. \ref{TDMA_PS_user_k}); such a joint CFO and CIR estimation method was proposed in \cite{RIS_pattern4} for a single-user scenario. As shown in Fig. \ref{TDMA_PS_user_k}, one pilot block is divided into $K$ non-overlapping time slots of length $N_{\rm{T}}=N/K$. It is assumed that  $N_{\rm{T}} \ge 2L$. The time slot $k$ is assigned to user $k$, and users transmit pilot sequences only in their assigned time slots (i.e., TDMA). For time slot $k$ in pilot block $b$, user $k$ transmits a length-$N_{\rm{T}}$ periodic pilot sequence of period $L$, i.e., $x_{k,b}(u+L)=x_{k,b}(u)$ for $u\in \{2(k-1)N_{\rm{T}}, 2(k-1)N_{\rm{T}}+1, \ldots, 2kN_{\rm{T}}-L-1 \}$. 

With this pilot structure, we can extend the joint CFO and CIR estimation method in \cite{RIS_pattern4} from the single-user to the multi-user scenario. For discrete time $u\in \{2(k-1)N_{\rm{T}}+L-1, 2(k-1)N_{\rm{T}}+L, \ldots, 2kN_{\rm{T}}-1 \}$ in pilot block $b$, the transmission of user $k$ does not exhibit any MUI when the length of each time slot is longer than $2L$. This means that the total overhead length is at least $2KL(R+1)$. Since the number of reflection elements $R$ is usually large even for a moderate size RIS, the resulting channel estimates can easily become outdated. Also, when $N$ is fixed, the system of \cite{RIS_pattern3} can support only half the number of users compared to the system of \cite{RIS_pattern4} (for \cite{RIS_pattern3}, $K_{\rm{max}} = N/L$, while for \cite{RIS_pattern4}, $K_{\rm{max}} = N/2L$). Consequently, a low pilot overhead joint CFO and CIR estimation method with high pilot resource efficiency is required for RIS-aided multi-user massive MIMO OFDM systems.

\section{Proposed Joint CFO and CIR estimation Method}

In this section, we propose a new joint CFO and channel estimation method for OFDM-based RIS-aided multi-user massive MIMO systems. First, the CFOs are estimated using a time-domain correlation-based approach. After compensating the estimated CFOs, the CIR matrix is estimated in the time domain.

\subsection{Proposed transmit sequence structure}

\begin{figure}[t]
    \centering
    \includegraphics[scale=0.35]{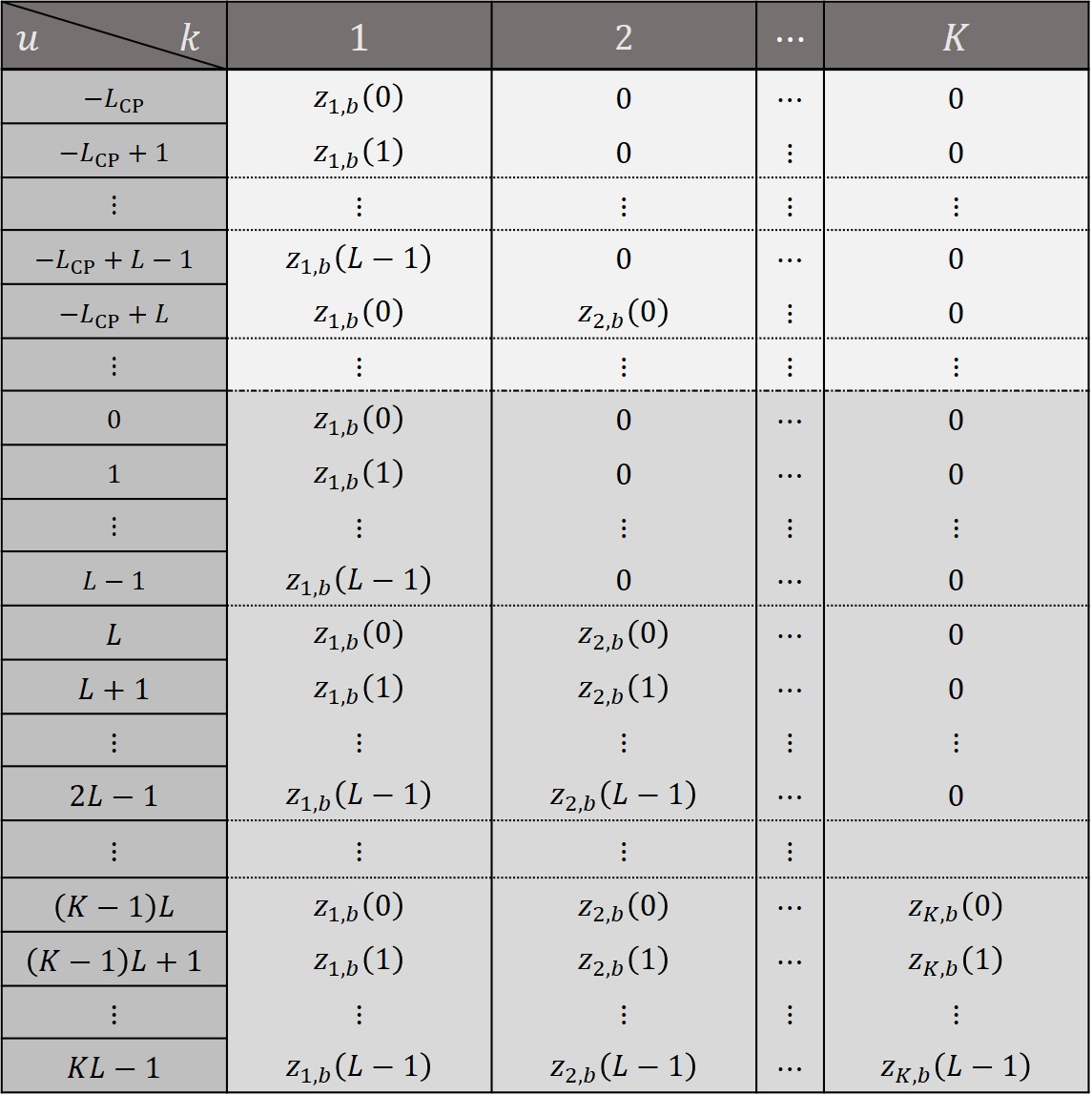}
    \caption{Proposed pilot structure for user $k$ in the time domain for pilot block $b$ with $((b))_K=0$, when $N=KL$.}
    \label{joint_PS_user_k}
\end{figure}

As shown in the previous section, the CFOs cannot be ignored when performing channel estimation. However, it is difficult to individually extract each CFO from the received sequences given in (\ref{eq:y_k_n}). It is relatively straightforward to extend the joint CFO and CIR estimation method proposed in \cite{RIS_pattern4} from the single-user to the multi-user case by using TDMA pilot sequences as shown in Fig. \ref{TDMA_PS_user_k}. The minimum number of pilots required for the TDMA-based CIR estimation method is $2KL(R+1)$, while the pilot overhead for the CIR estimation method using OFDMA in \cite{RIS_pattern4} is only $KL(R+1)$. However, the OFDMA-based method in \cite{RIS_pattern4} requires extra pilot resources for CFO estimation. Because the length of the overhead for channel estimation changes linearly with the number of RIS elements (which in turn is usually large in practical applications), it is not desirable to use more resources than $KL(R+1)$. 

In the following, we propose a pilot sequence structure which requires $KL(R+1)$ pilot resources as shown in Fig. \ref{joint_PS_user_k}. Using this structure, the same pilot resources can be used for both CFO and CIR estimation, and in contrast to \cite{RIS_pattern3}, no extra overhead is required. For example, an extra pilot overhead of length $2L(R+1)$ is required for the method of \cite{RIS_pattern3} to obtain the same number of correlation samples as are used for the proposed method.

The pilot symbol structure consists of $R+1$ symbol blocks of length $N$. The pilot sequence is divided into $N_{\rm{T}}=N/L \in \mathbb{Z}$ time slots. For simplicity, we assume that $N_{\rm{T}} = K$ and that $(R+1)/K\in \mathbb{Z}$. As shown in Fig. \ref{joint_PS_user_k}, for pilot block $b$, user $k$ transmits a sequence of length $N$ which consists of a sequence of $L( k-((b))_K-1 )$ zeros followed by a periodic sequence of period $L$, i.e.,
\begin{equation}\label{eq::pilot_b01}
    \boldsymbol{x}_{k,b}=
        \left[
            \boldsymbol{0}_{1 \times L( k-((b))_K-1 )},\;
            \boldsymbol{z}_{k,b}^{\rm{T}}
        \right]
        ^{\rm{T}}
        .
\end{equation}
The periodic part of the pilot sequence for one user consists of a length-$L$ Zadoff-Chu (ZC) sequence, which is different from the ZC sequences for other users.


\subsection{Correlation-based CFO estimation}

Because of the CP design, $\tilde{y}_{k,m,b}(u_2) = \tilde{y}_{k,m,b}(u_1)$ where $u_1=L-1$ and $u_2=L-1-L_{\rm{CP}}$. Based on the pilot sequence design (\ref{eq::pilot_b01}), the received signal (\ref{eq:y_k_n}) at discrete time instant $u_1$ for pilot block $b$ with $((b))_K=k-1$ (i.e., pilot block assigned for user $k$) can be rewritten as
\begin{equation}\label{eq:y_mb_u_1}
    y_{m,b}(u_1)=
        e^{j\frac{2\pi \epsilon_{k} 
            (b L_{\rm{s}} + u_1 )} {N}}
            \widetilde{y}_{k,m,b}(u_1)
        +v_{m,b}(u_1)
        = y_{k,m,b}(u_1)
        .
\end{equation}
Due to the CP design, the corresponding CP part of the received signal at time instant $u_2$ can be given as
\begin{equation}\label{eq:y_mb_u_2}
    y_{m,b}(u_2)=
        e^{j\frac{2\pi \epsilon_{k} 
            (b L_{\rm{s}} + u_2 )} {N}}
            \widetilde{y}_{k,m,b}(u_1)
        +v_{m,b}(u_2)
        = y_{k,m,b}(u_2)
        .
\end{equation}
Consequently, the correlation between $y_{k,m,b}(u_1)$ and $y_{k,m,b}(u_2)$ can be obtained as
\begin{equation}\label{eq::c_m(kappa)}
\begin{split}
    c_{k,m,b}=&
        y_{k,m,b}^*(u_2) y_{k,m,b}(u_1)
        \\
        =&
        e^{j\frac{2\pi \epsilon_{t} L_{\rm{CP}} }
        {N}}
        \left\vert
            \widetilde{y}_{k,m,b}(u_1)
        \right\vert^2
        +\nu_{k,m,b}(u_1)
        ,
\end{split}
\end{equation}
where $
    \nu_{k,m,b}(u_1)=
        v_{m,b}^*(u_2)
        e^{j\frac{2\pi \epsilon_{k} ( b L_{\rm{s}} + u_1)}{N}}
        \widetilde{y}_{k,m,b}(u_1) 
        +
        e^{-j\frac{2\pi \epsilon_{k} (b L_{\rm{s}}+ u_2)}{N}}
        \widetilde{y}_{k,m,b}^*(u_2) 
        v_{m,b}(u_1) 
        +
        v_{m,b}^*(u_2) 
        v_{m,b}(u_1)
        $.

Since $x_{k,b}(u)$, $g_{k,m,r}(l)$ and $v_{m,b}(u)$ are independent random variables, we have $\mathbb{E} \{ v_{m,b}(u) \} = \mathbb{E} \{ \nu_{k,m,b}(u_1) \} = 0$. Therefore, we can mitigate the influence of the noise term $\nu_{k,m,b}(L-1)$ by averaging the correlation samples over $(R+1)/K$ pilot blocks (since $(R+1)/K\in\mathbb{Z}$) and $M$ BS antennas. After sharing the received signals between BS antennas, the CFO of user $k$ can be estimated as
\begin{equation}\label{eq::c_(kappa)}
\begin{split}
    \hat{\epsilon}_{k}
        =&
        \frac{N
        \angle\left(
            \frac{K}{M (R+1)} 
            \sum_{m=1}^{M}\sum_{b=1}^{(R+1)/K} c_{k,m,(b-1)K+k-1}
        \right)
        }
        {2 \pi L} 
        \\
        = &
        \frac{N
        \angle\left(
            e^{j\frac{2\pi \epsilon_{t} L }{N}}
            \check{y}_{k}(u_1)
            +
            \bar{\nu}_{k}(u_1)
        \right)
        }
        {2 \pi L} 
        ,
\end{split}
\end{equation}
where $\check{y}_{k}(u_1) = \sum_{m=1}^{M} \sum_{b=1}^{(R+1)/K} K \left\vert \widetilde{y}_{k,m,(b-1)K+k-1}(u_1) \right\vert^2$ $/(M(R+1))$, and $\bar{\nu}_{k}(u_1)= \sum_{m=1}^{M}$ $\sum_{b=1}^{(R+1)/K}
K \nu_{k,m,(b-1)K+k-1}(u_1) / (M(R+1))$. 

The expected value of $\bar{\nu}_{k} (u_1)$ and the variance of $\bar{\nu}_{k}(u_1)$ can be derived as
\begin{equation*}
\begin{split}
    \mathbb{E} [ \bar{\nu}_{k}(u_1) ] 
    =&
    \frac{ \sum_{m=1}^{M}         
           \sum_{b=1}^{(R+1)/K}
                 K \mathbb{E} \left[\nu_{k,m,(b-1)K+k-1}(u_1) \right]}
                 {M(R+1)}  
    \\
    =&
    0  
    ,
\end{split}
\end{equation*}
and
\begin{equation*}
\begin{split}
    \text{Var} ( \bar{\nu}_{k}(u_1) )
    =& \;
    \sum\limits_{m=1}^{M}
    \sum\limits_{l=0}^{L-1} 
    \sum\limits_{r=0}^{R} 
    \frac{ 2 K \sigma^2 
           \tilde{\sigma}_{k,m,r,l}^2 }
         {M^2(R+1)} 
    +
    \frac{ K \sigma^4}
         {M(R+1)} 
    \\
    =&
    \frac{ 2 K \sigma^2 
           \tilde{\sigma}_{k}^2 }
         {M} 
    +
    \frac{ K \sigma^4}
         {M(R+1)} 
    ,
\end{split}
\end{equation*}
respectively, where $\tilde{\sigma}_{k,m,r,l}^2$ is Var$(g_{k,m,r}(l))$ (since the BS antennas are collocated and the same is true for the RIS elements, we assume that $\tilde{\sigma}_{k,m,r,l}^2=\tilde{\sigma}_{k,l}^2$ and $\tilde{\sigma}_{k}^2=\sum_{l=0}^{L-1} \tilde{\sigma}_{k,l}^2$). Consequently, when $M$ and $R$ are large, the noise term becomes sufficiently small due to the averaging.

\subsection{Least-squares CIR estimation in time domain}

For time slot $t \in \{ 1, 2, \ldots, K \}$, the time-domain received signal at discrete time $u$ is comprised of a superposition of signals from $t$ users. Consequently, by time slot $t$ in pilot block $b$, the received signal vector at BS antenna $m$ can be written as
\begin{equation}\label{eq:y_mb_u_t}
    \boldsymbol{y}_{m,t,b}=
        \boldsymbol{\mathcal{X}}_{t,b}
        \boldsymbol{\bar{g}}_{m,t,b}
        +
        \boldsymbol{v}_{m,t,b}
        ,
\end{equation}
where $\boldsymbol{y}_{m,t,b}= [ \boldsymbol{y}_{m,b}(0), \boldsymbol{y}_{m,b}(1), \ldots, \boldsymbol{y}_{m,b}(tL-1) ]^{\rm{T}}$, $\boldsymbol{v}_{m,t,b}=[ v_{m,b}(0), v_{m,b}(1), \ldots, v_{m,b}(tL-1) ]^{\rm{T}}$, $\boldsymbol{\bar{g}}_{m,t,b} = [ \boldsymbol{\bar{g}}_{1,m,b}, \boldsymbol{\bar{g}}_{2,m,b}, \ldots, \boldsymbol{\bar{g}}_{t,m,b} ]^{\rm{T}}$,  $\boldsymbol{\bar{g}}_{k,m,b} = [ \bar{g}_{k,m,b}(0), \bar{g}_{k,m,b}(1), \ldots, \bar{g}_{k,m,b}(L-1) ]^{\rm{T}}$,

\begin{equation*}
    \boldsymbol{\mathcal{X}}_{t,b}=
    \begin{bmatrix}
            \boldsymbol{\tilde{x}}_{1,b}(0) & \boldsymbol{\tilde{x}}_{2,b}(0) & \cdots & \boldsymbol{\tilde{x}}_{t,b}(0)
            \\
            \boldsymbol{\tilde{x}}_{1,b}(1) & \boldsymbol{\tilde{x}}_{2,b}(1) & \cdots & \boldsymbol{\tilde{x}}_{t,b}(1)
            \\
            \vdots  &   \vdots  &   \ddots  &   \vdots
            \\
            \boldsymbol{\tilde{x}}_{1,b}(tL-1) & \boldsymbol{\tilde{x}}_{2,b}(tL-1) & \cdots & \boldsymbol{\tilde{x}}_{t,b}(tL-1)
        \end{bmatrix}
    ,
\end{equation*}
$\boldsymbol{\tilde{x}}_{k,b}(u) = e^{j\frac{2\pi\epsilon_{k} (b L_{\rm{s}}+u)} {N}} \boldsymbol{x}_{k,b}(u)$, and $\boldsymbol{x}_{k,b}(u)=[x_{k,b}(u), x_{k,b}(u-1), \ldots, x_{k,b}(u-L+1)]$.
Since $\boldsymbol{\mathcal{X}}_{t,b}$ consists of known parameters and CFO estimates, we can obtain its estimate $\widehat{\boldsymbol{\mathcal{X}}}_{t,b}$. $\widehat{\boldsymbol{\mathcal{X}}}_{t,b}$ can be guaranteed to be invertible through appropriate choice of the sequences $\boldsymbol{x}_{k,b}(u)$.

Assuming perfect CFO estimation ($\hat{\epsilon}_{k}=\epsilon_{k}$ for all $k$), we can estimate the channel vector $\boldsymbol{\bar{g}}_{m,t,b}$ as
\begin{equation}\label{eq:g_kmb_hat1}
    \boldsymbol{\hat{\bar{g}}}_{m,t,b}
        =
        \widehat{\boldsymbol{\mathcal{X}}}_{t,b}^{-1}
        \boldsymbol{y}_{m,t,b}
        =
        \boldsymbol{\bar{g}}_{m,t,b}
        +
        \boldsymbol{\tilde{v}}_{m,t,b}
        ,
\end{equation}
where $\boldsymbol{\tilde{v}}_{m,t,b}=\widehat{\boldsymbol{\mathcal{X}}}_{t,b}^{-1}\boldsymbol{v}_{m,t,b}$.

After repeating (\ref{eq:g_kmb_hat1}) over all $t$, the CIR vector for user $k$ can be obtained as
\begin{equation}\label{eq:g_kmb_hat2}
    \boldsymbol{\hat{\bar{g}}}_{k,m,b}
        =
        \frac{1}{N_{\rm{T},k,b}}
        \sum_{t=1}^{N_{\text{T},k,b}}
        \boldsymbol{\hat{\bar{g}}}_{k,m,t,b}
        =
        \boldsymbol{\bar{g}}_{k,m,b}
        +
        \boldsymbol{\bar{v}}_{k,m,b}
        ,
\end{equation}
where $N_{\text{T},k,b}=N_{\text{T}}-k+((b))_K+1$, $\boldsymbol{\bar{v}}_{k,m,b}=\frac{1}{N_{\text{T},k,b}} \sum_{t=1}^{N_{\text{T},k,b}} \boldsymbol{v}_{k,m,t,b}$ $\boldsymbol{\hat{\bar{g}}}_{m,t,b}=[\boldsymbol{\hat{\bar{g}}}_{1,m,t,b}, \boldsymbol{\hat{\bar{g}}}_{2,m,t,b}, \ldots,$ $ \boldsymbol{\hat{\bar{g}}}_{t,m,t,b} ]^{\rm{T}}$, 
$\boldsymbol{\hat{\bar{g}}}_{k,m,t,b}=[\hat{\bar{g}}_{m,t,b}((k-1)L), \hat{\bar{g}}_{m,t,b}((k-1)L+1), \ldots, \hat{\bar{g}}_{m,t,b}(kL-1) ]^{\rm{T}}$, 
$\boldsymbol{v}_{m,t,b}=[ \boldsymbol{v}_{1,m,t,b},$ $ \boldsymbol{v}_{2,m,t,b}, \ldots, \boldsymbol{v}_{t,m,t,b} ]^{\rm{T}}$
, and 
$\boldsymbol{v}_{k,m,t,b}= [ \boldsymbol{\tilde{v}}_{m,t,b}((k-1)L), \boldsymbol{\tilde{v}}_{m,t,b}((k-1)L+1), \ldots, \boldsymbol{\tilde{v}}_{m,t,b}(kL-1)]$.

By stacking $\hat{\bar{\boldsymbol{g}}}_{k,m,b}$ over $b$, we can obtain 
\begin{equation}\label{eq:G_km_hat1}
    \widehat{\boldsymbol{\bar{G}}}_{k,m}
        =
        [
            \boldsymbol{\bar{g}}_{k,m,0},
            \boldsymbol{\bar{g}}_{k,m,1},
            \ldots ,
            \boldsymbol{\bar{g}}_{k,m,R} 
        ]
        =
        \boldsymbol{G}_{k,m}
        \boldsymbol{\Phi}
        +\boldsymbol{\bar{V}}_{k,m}
        ,
\end{equation}
where $\boldsymbol{G}_{k,m}= [ \boldsymbol{g}_{k,m,0}, \boldsymbol{g}_{k,m,1}, \ldots, \boldsymbol{g}_{k,m,R} ]$, and $\boldsymbol{\bar{V}}_{k,m}=[ \boldsymbol{\bar{v}}_{k,m,0}, \boldsymbol{\bar{v}}_{k,m,1}, \ldots, \boldsymbol{\bar{v}}_{k,m,R}]$.

Using the unitary property of $\boldsymbol{\Phi}$, we can estimate the channel matrix $\boldsymbol{G}_{k,m}$ for BS antenna $m$ as
\begin{equation}\label{eq:G_km_hat2}
    \boldsymbol{\widehat{G}}_{k,m}
        =
        \widehat{\boldsymbol{\bar{G}}}_{k,m}
        \boldsymbol{\Phi}^{-1}
        =
        \boldsymbol{G}_{k,m}
        +
        \boldsymbol{\bar{V}}_{k,m}
        \boldsymbol{\Phi}^{-1}
        .
\end{equation}

\section{Computational complexity analysis and comparison}

In this section, we analyze and compare the computational complexity of our proposed joint CFO and CIR estimation method with the approach in \cite{RIS_pattern4} and the CIR estimation method in \cite{RIS_pattern3} in terms of the number of complex multiplications.

The total computational complexity of the proposed method is given by
\begin{equation}\label{eq::C_pro}
\begin{split}
    C_{\rm{Pro}}
    =&
    C_{\rm{CFO,Pro}}
    +
    C_{\rm{CIR,Pro}}
    \\
    =&
    \mathcal{O}
    (
        R^3
        +
        NMR^2
        +
        N^2MR
        +
        N^3M
    )
    ,
\end{split}
\end{equation}
where $C_{\rm{CFO,Pro}} = \mathcal{O}(MR)$ is the computational complexity of the proposed CFO estimation method (evaluation of (\ref{eq::c_(kappa)}), and  $C_{\rm{CIR,Pro}} = \mathcal{O} (R^3+NMR^2+N^2MR+N^3M)$ is the computational complexity of the proposed CIR estimation method (evaluation of (\ref{eq:g_kmb_hat1})-(\ref{eq:G_km_hat2})).

The computational complexity of the method of \cite{RIS_pattern3} (evaluation of (\ref{eq:hat_G_km_no_CFO})) is given by
\begin{equation}\label{eq::C_S1}
    C_{\rm{CIR,S1}}
    =
    \mathcal{O}
    (
        R^3
        +
        NMR^2
        +
        N^2MR
    )
    .
\end{equation}
The total computational complexity of the method in \cite{RIS_pattern4} is given by
\begin{equation}\label{eq::C_S2}
    C_{\rm{S2}}
    =
    C_{\rm{CFO,S2}}
    +
    C_{\rm{CIR,S2}}
    =
    \mathcal{O}
    (
        R^3
        +
        NMR^2
        +
        LNMR
    )
    ,
\end{equation}
where $C_{\rm{CFO,B1}}=\mathcal{O}(KMR)$ and $C_{\rm{CIR,B1}} = \mathcal{O}(R^3+NMR^2+LNMR)$ are the computational complexity of CFO estimation and CIR estimation, respectively. 

\begin{figure}[t]
    \centering
    \includegraphics[scale=0.5]{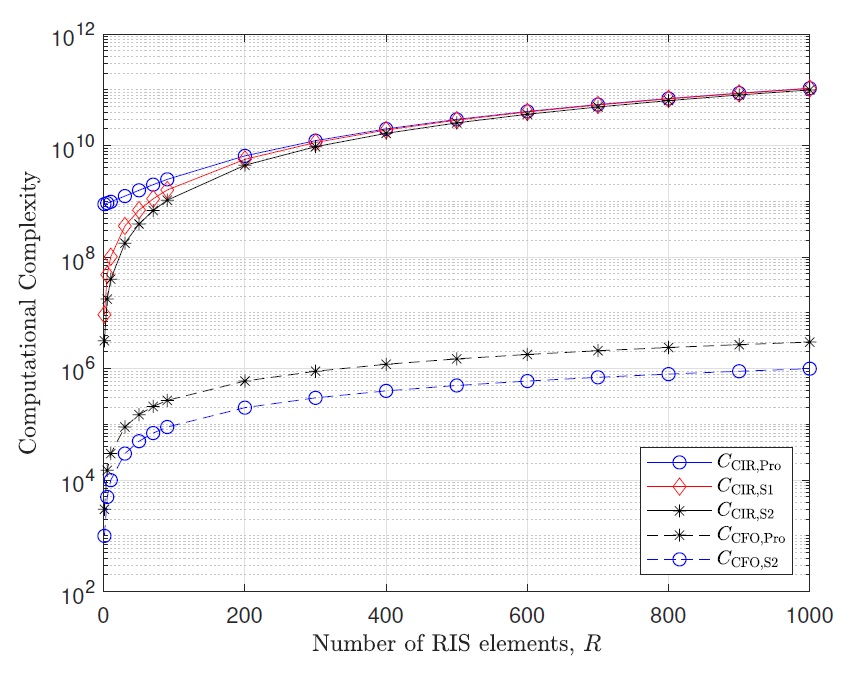}
    \caption{Computational complexity as a function of $R$ when $K=3$, $L=32$, and $M=1000$.}
    \label{Complexity}
\end{figure}

Fig. \ref{Complexity} shows the numerical evaluation of the computational complexity of the proposed method and of the (appropriately enhanced) approachs in \cite{RIS_pattern3, RIS_pattern4}. As shown in (\ref{eq::C_pro}) and (\ref{eq::C_S2}), in the computational complexity of joint CFO and CIR estimation, the computational complexity of CIR estimation is much higher than that of correlation-based CFO estimation. When $N$ is a constant and $R$ increases, the computational load for all the methods under study become close to one another, and they converge when $R\gg N$ as shown in Fig. \ref{Complexity} (when $R\gg N$, $C_{\rm{Pro}} \approx C_{\rm{S1}} \approx C_{\rm{S2}} \approx \mathcal{O} (R^3+NMR^2)$).

\section{RIS reflection coefficient optimization}

In the previous section, it has been demonstrated that the computational complexity of joint CFO and CIR estimation is proportional to $R$. Similarly, the complexity of optimizing the RIS phase shifts increases rapidly with $R$, especially when a grid search is utilized. Since $R$ is in general large for RIS-aided systems, the optimization of the RIS phase shifts may require a long time and in some cases can even exceeded the coherence time of the channel. Consequently, for RIS-aided systems, it is important to reduce the computational complexity and the execution time, not only for CFO/CIR estimation but also for RIS optimization. Motivated by this, we propose a low-complexity line search method with projected gradients using an adaptive step size for optimizing the RIS reflection coefficients.

\subsection{Problem statement}

After estimating the CFO and CIR in the training phase, the RIS reflection coefficients for the data transmission phase need to be optimized in order to maximize the achievable rate in the considered communication system. This problem can be formulated as

\begin{equation}
    \text{(P1):} \;
    \max_{\boldsymbol{\phi}_{\rm{d}}}
    f_1(\boldsymbol{\phi}_{\rm{d}})
    =
    \frac{
    \sum\limits_{m=1}^{M}
    \sum\limits_{N=0}^{N-1}
        \log_{2}
        \left(
            1+
            \frac{P
            \sum_{k=1}^{K}
            \widehat{\mathcal{G}}_{k,m,n}(\boldsymbol{\phi}_{\rm{d}})}
            {N \Upsilon \sigma^2}
        \right)}
    {N+L_{\rm{CP}}}
\end{equation}
\begin{equation}
    \text{Subject to } \; \vert \phi_{\rm{d},r} \vert =1 \; \forall \; r=1,2, \ldots, R,
\end{equation}
where $\widehat{\mathcal{G}}_{k,m,n}(\boldsymbol{\phi}_{\rm{d}})=\vert \sum_{r=0}^{R} \gamma_{k}(n)\hat{h}_{k,m,r}(n)\phi_{\rm{d},r} \vert^2$ is the estimated frequency-domain channel gain between user $k$ and BS antenna $m$ on subcarrier $n$, which varies with the vector of reflection coefficients $\boldsymbol{\phi}_{\rm{d}}=[1, \phi_{\rm{d},1}, \ldots, \phi_{\rm{d},R}]^{\rm{T}}$, and $\Upsilon\ge 1$ denotes the achievable rate gap due to the use of practical modulation and coding schemes. 

\subsection{Projected gradient method}

Removing the constant denominator, the optimization problem (P1) can be rewritten as
\begin{equation}\label{eq::P2}
    \text{(P2):} \;
    \max_{\boldsymbol{\phi}_{\rm{d}}}
    f_2(\boldsymbol{\phi}_{\rm{d}})
    =
    \sum\limits_{m=1}^{M}
    \sum\limits_{N=0}^{N-1}
        \ln
        \left(
            1+
            \frac{P
            \sum_{k=1}^{K}
            \widehat{\mathcal{G}}_{k,m,n}(\boldsymbol{\phi}_{\rm{d}})}
            {N \Upsilon \sigma^2}
        \right)
\end{equation}
\begin{equation}
    \text{Subject to } \; \vert \phi_{\rm{d},r} \vert =1 \; \forall \; r=1,2, \ldots, R.
\end{equation}
After differentiating $f_2(\boldsymbol{\phi}_{\rm{d}})$ with respect to $\boldsymbol{\phi}_{\rm{d},\tilde{r}}^*$ for $\tilde{r}\in\{1,2,\ldots,R\}$, we obtain\footnote{$a^*$ denotes complex conjugate of a complex number $a$.}
\begin{equation}\label{eq::diff_f2}
    \frac{d}{d\phi_{\rm{d},\tilde{r}}^*}
    f_2(\boldsymbol{\phi}_{\rm{d}})
    =
    \sum\limits_{m=1}^{M}
    \sum\limits_{N=0}^{N-1}
        A_{m,n}^{-1}
        \frac{P}{N \Upsilon \sigma^2}
        \sum_{k=1}^{K}
        \frac{d}{d\phi_{\rm{d},\tilde{r}}^*}
        \widehat{\mathcal{G}}_{k,m,n}(\boldsymbol{\phi}_{\rm{d}})
    ,
\end{equation}
where $A_{m,n}=1 + P \sum_{k=1}^{K} \widehat{\mathcal{G}}_{k,m,n}(\boldsymbol{\phi}_{\rm{d}})/N \Upsilon \sigma^2$. Furthermore, we have
\begin{equation}\label{eq::diff_f2_r}
\begin{split}
    \frac{d}{d\phi_{\rm{d},\tilde{r}}^*} &
        \widehat{\mathcal{G}}_{k,m,n}(\boldsymbol{\phi}_{\rm{d}})
    = \;
    \frac{d}{d\phi_{\rm{d},\tilde{r}}^*}
    \left\vert
        \sum_{r=0}^{R}
            \gamma_{k}(n)
            \hat{h}_{k,m,r}(n)\phi_{d,r}
    \right\vert^2
    \\
    =&\;
    \vert
        \gamma_{k}(n)
    \vert^2
    \frac{d}{d\phi_{\rm{d},\tilde{r}}^*}
    \left\vert
        \sum_{r=0}^{R}
            \hat{h}_{k,m,r}(n)\phi_{d,r}
    \right\vert^2
    \\
    =&\;
    \vert
        \gamma_{k}(n)
    \vert^2
    \frac{d}{d\phi_{\rm{d},\tilde{r}}^*}
        \sum_{r_1=0}^{R}
            \hat{h}_{k,m,r_1}(n)\phi_{d,r_1}
        \sum_{r_2=0}^{R}
            \hat{h}_{k,m,r_2}^*(n)\phi_{d,r_2}^*
    \\
    =&\;
    \vert
        \gamma_{k}(n)
    \vert^2
    \sum_{r_1=0}^{R}
        \hat{h}_{k,m,r_1}(n)\phi_{d,r_1}
    \frac{d}{d\phi_{\rm{d},\tilde{r}}^*}
    \sum_{r_2=0}^{R}
        \hat{h}_{k,m,r_2}^*(n)\phi_{d,r_2}^*
    \\
    =&\;
    \vert
        \gamma_{k}(n)
    \vert^2
    \hat{h}_{k,m,\tilde{r}}^*(n)
    \sum_{r=0}^{R}
        \hat{h}_{k,m,r}(n)\phi_{d,r}
    .
\end{split}
\end{equation}

Substituting (\ref{eq::diff_f2_r}) into (\ref{eq::diff_f2}), we obtain
\begin{equation}\label{eq::diff_f2_final}
\begin{split}
    \frac{d}{d\phi_{\rm{d},\tilde{r}}^*}
    f_2(\boldsymbol{\phi}_{\rm{d}})
    =& 
    \frac{P}{N \Upsilon \sigma^2}
    \sum\limits_{m=1}^{M}
    \sum\limits_{N=0}^{N-1}
        A_{m,n}^{-1}
        \sum_{k=1}^{K}
        \vert
            \gamma_{k}(n)
        \vert^2
        \\
        &\times
        \sum_{r=0}^{R}
            \hat{h}_{k,m,r}(n)\phi_{d,r}
        \hat{h}_{k,m,\tilde{r}}^*(n)
    .
\end{split}
\end{equation}

The gradient of $f_2(\boldsymbol{\phi}_{\rm{d}})$ with respect to $\boldsymbol{\phi}_{\rm{d}}^*$ can be written as
\begin{equation}\label{eq::gradient_P2}
\begin{split}
    \nabla_{\boldsymbol{\phi}_{\rm{d}}^*}&
    f_2(\boldsymbol{\phi}_{\rm{d}})
    =
    \left[
        \frac{d f_2(\boldsymbol{\phi}_{\rm{d}})}{d \phi_{\rm{d},1}^*}
        ,
        \frac{d f_2(\boldsymbol{\phi}_{\rm{d}})}{d \phi_{\rm{d},2}^*}
        ,
        \cdots
        ,
        \frac{d f_2(\boldsymbol{\phi}_{\rm{d}})}{d \phi_{\rm{d},R}^*}
    \right]^{\rm{T}}
    \\
    =&
    \frac{P}{N \Upsilon \sigma^2}
    \left[
        \hat{h}_{k,m,1}^*(n)
        ,
        \hat{h}_{k,m,2}^*(n)
        ,
        \cdots
        ,
        \hat{h}_{k,m,R}^*(n)
    \right]^{\rm{T}}
    \\
    &\times
    \sum\limits_{m=1}^{M}
    \sum\limits_{N=0}^{N-1}
        A_{m,n}^{-1}
        \sum_{k=1}^{K}
        \vert
            \gamma_{k}(n)
        \vert^2
        \sum_{r=0}^{R}
            \hat{h}_{k,m,r}(n)\phi_{d,r}.
\end{split}
\end{equation}

After computing the gradient above, we update the value of $\boldsymbol{\phi}_{\rm{d}}$ in iteration $\tau$ according to
\begin{equation}\label{eq::gradient2}
    \boldsymbol{\phi}_{\rm{d}}^{(\tau+1)}
    =
    P\left(
        \boldsymbol{\phi}_{\rm{d}}^{(\tau)}
        +
        \mu
        \nabla_{\boldsymbol{\phi}_{\rm{d}}^*} f_2(\boldsymbol{\phi}_{\rm{d}}^{(\tau)})
    \right)
    ,
\end{equation}
where $\tau \in \{1, 2, \ldots, N_{\tau}\}$ and $\mu$ are the iteration index and the step size, respectively, $N_{\tau}$ is the number of iterations, and $P(\cdot)$ is the projection operator. The gradient projection $\tilde{\phi}_{\rm{d},r}^{(\tau)} = P \left( \phi_{\rm{d},r}^{(\tau)} \right)$ is given by
\begin{equation}
    \tilde{\phi}_{\rm{d},r}^{(\tau)}=
        \begin{cases}
            \phi_{\rm{d},r}^{(\tau)}
            /
            \vert \phi_{\rm{d},r}^{(\tau)} \vert
            ,
            & \text{for} \; \phi_{\rm{d},r}^{(\tau)}\neq 0 
            \\
            e^{j\alpha},
            & \text{for} \; \phi_{\rm{d},r}^{(\tau)} = 0
        \end{cases}
        ,
\end{equation}
where $\alpha\in[0, 2\pi]$. Consequently, we simultaneously optimize all the RIS reflection coefficients in each iteration of the proposed algorithm.

In order to make the proposed algorithm computationally and time efficient, we use a line search procedure to adjust the step size $\mu$. The step size $\mu$ in (\ref{eq::gradient2}) can be found as $\mu_{0} \varrho^{k_{\tau}}$, where $k_{\tau}$ is the smallest non-negative integer such that
$
    f_2 \left( \boldsymbol{\phi}_{\rm{d}}^{(\tau+1)} \right) 
    - 
    f_2 \left( \boldsymbol{\phi}_{\rm{d}}^{(\tau)} \right)
    \ge
    \delta_{\phi}
    \left\Vert
        \boldsymbol{\phi}_{\rm{d}}^{(\tau+1)}
        -
        \boldsymbol{\phi}_{\rm{d}}^{(\tau)}
    \right\Vert^2
    ,
$ where $\mu_0 > 0$ is the initial value of $\mu$, $\delta_{\phi} > 0$ is a small constant, and $\varrho\in (0,1)$ \cite{COR_RIS}. The proposed PGM procedure ensures that the objective sequence increases after each iteration. Thus, the PGM is guaranteed to converge to a stationary point of (\ref{eq::P2}), which is, however, not necessarily a globally optimal solution. Consequently, we can optimize the RIS reflection coefficient vector $\boldsymbol{\phi}_{\rm{d}}$ using the proposed PGM with a low complexity.

\section{Numerical Results and Discussion}

\begin{figure}[t]
    \centering
    \includegraphics[scale=0.27]{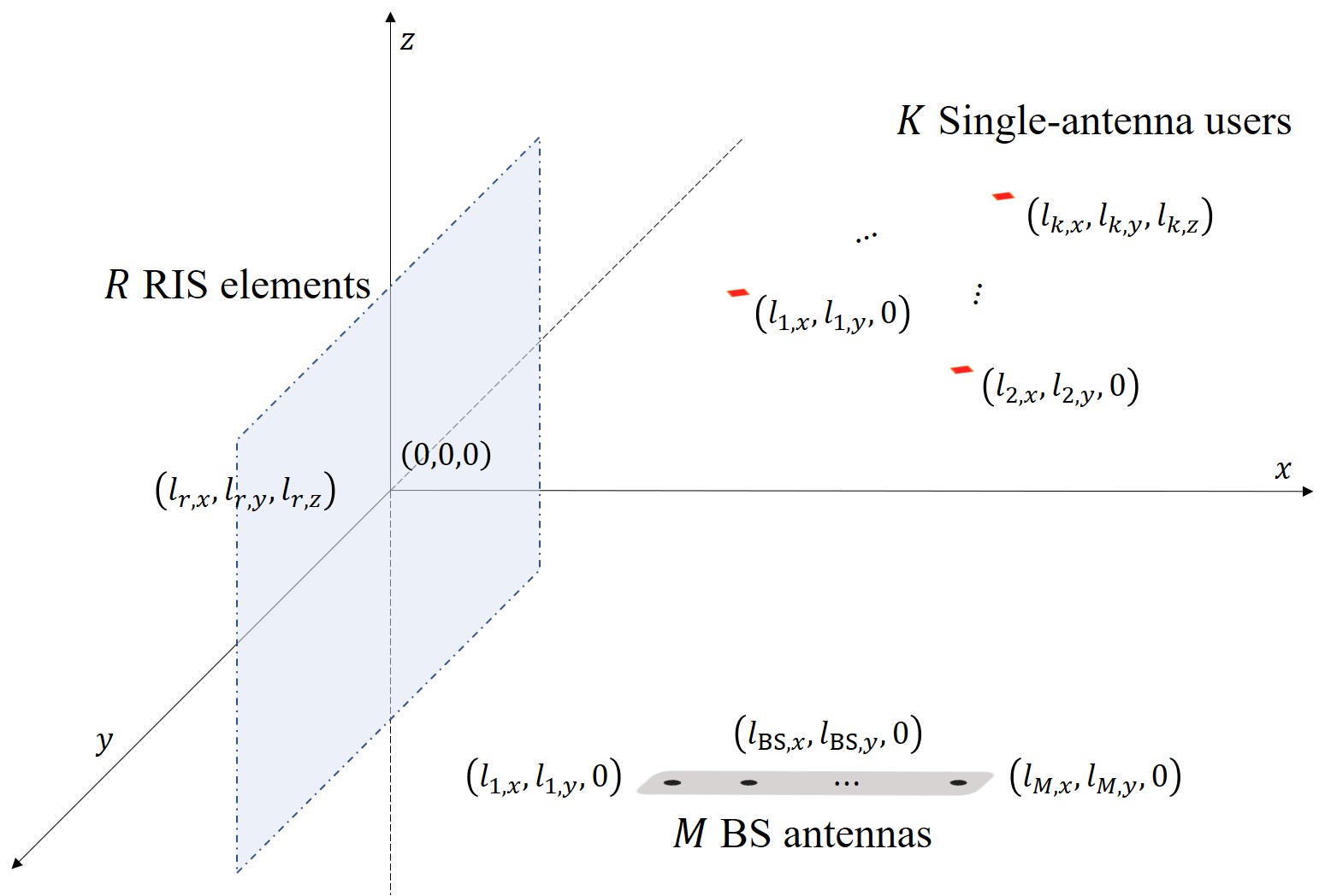}
    \caption{3D representation of the BS antennas, the users and the RIS.}
    \label{dep_diagram}
\end{figure}

In this section, we evaluate the performance of the proposed joint CFO and CIR estimation method. As shown in Fig. \ref{dep_diagram}, $M$ BS antennas, $K$ single-antenna users, and $R$ RIS elements are considered in a three-dimensional (3D) Cartesian coordinate system. The RIS is located in the $yz$-plane and the position of its midpoint is $(0, 0, 0)$. We assume that the RIS elements are placed in a uniform rectangular array (URA). The distance between the centers of adjacent RIS elements in both $y$- and $z$-dimensions is $\Delta_{\rm{RIS}}$. The BS antennas are placed in a uniform linear array (ULA). For simplicity, we assume here that the BS antennas are located in the $xy$-plane, where $x, y \ge 0$, and its ULA is parallel to the $x$-axis. The position of its midpoint is set as $(l_{\rm{BS},x}, l_{\rm{BS},y}, 0)$. The distance between the adjacent BS antennas is $\Delta_{\rm{BS}}$. We consider $\Delta_{\rm{BS}}$ to be equal to $\lambda/2$ to avoid antenna coupling issues or correlated channels for different antennas, where $\lambda$ denotes the wavelength. For simplicity, we assume that users are also located also in the $xy$-plane, where $x \ge 0$ and $y \le 0$. The position of user $k$ is set as $(l_{k,x}, l_{k,y}, 0)$ and we assume that the users' positions are fixed while the CFO and CIR are being estimated. 
In the following simulations, all of the CIR vectors are modeled according to a  frequency-selective fading channel model with a delay spread of $L$. The first tap is set as a deterministic channel component and the remaining taps follow the Rayleigh channel distribution. The value $\kappa$ is defined as the ratio of the signal power in the dominant first tap channel component over the total scattered power in the remaining channel components. In this paper, we set $\kappa=4$ dB. 

The (far-field) free space path loss (FSPL)\footnote{The FPSL is defined with respect to the midpoint of the RIS and that of the BS antenna array. The reason is that the distances between neighboring BS antennas are very small compared to the propagation distances. The same is true for the distances between the neighboring RIS elements.} for the link from user $k$ to the BS is equal to $\rho_{k}  = \frac{256 \pi^2  \tilde{d}_{\text{UR},k}^2 \tilde{d}_{\text{RB}}^2} {\mathcal{G}_{\rm{t}} \mathcal{G}_{\rm{r}} \lambda^4 \cos{\theta_{\text{T},k}} \cos{\theta_{\rm{R}}}}$ (see Eq. (7) and (9) in \cite{channel_model2}), and $\mathcal{G}_{\rm{t}}$ and $\mathcal{G}_{\rm{r}}$ are the transmit and receive antenna gains, respectively. The values of $\mathcal{G}_{\rm{t}}$ and $\mathcal{G}_{\rm{r}}$ are set to 2, since we assume that these antennas radiate/sense signals to/from the relevant half space \cite{channel_model2}. $\tilde{d}_{\text{UR},k}$ and $\tilde{d}_{\rm{RB}}$ are the distance between the antenna of user $k$ and the midpoint of the RIS and the distance between the midpoint of the RIS and the midpoint of the BS. $\theta_{\text{T},k}$ and $\theta_{\rm{R}}$ are the angle between the incident wave propagation direction from user $k$ and the normal to the midpoint of the RIS and the angle between the normal to the midpoint of the RIS and the reflected wave propagation direction to the midpoint of the BS, respectively. We neglect the spatial correlation among the elements of a CIR vector from user $k$ to RIS element $r$ and a CIR vector from RIS element $r$ to BS antenna $m$ for all $k$, $r$ and $m$. 

As for the simulation setup, we set $f = 2$ GHz (i.e., $\lambda = 15$ cm), $\Delta_{\rm{RIS}} = \Delta_{\rm{BS}} = \lambda/2=7.5$ cm, $(l_{\rm{BS},x}, l_{\rm{BS},y}, 0)=(200 \; \text{m}, 200 \; \text{m}, 0)$, $(l_{k,x}, l_{k,y}, 0)=(200+k-1 \; \text{m}, -200 \; \text{m}, 0)$, $\kappa=4$ dB, $K = 3$, $L = 32$, $L_{\rm{CP}}=L+2=34$, $\mu_{0}=1000$, $\delta_{\phi}=10^{-5}$, and $\varrho=0.5$. CFOs are generated from a uniform distribution within the range $(-0.5,0.5]$. The results in all figures are averaged over 5000 independent realizations of the channels and CFOs. We provide the NMSE performance, which is given by $\eta_{\epsilon} = \frac{1}{K} \mathbb{E}\left\{ \frac{\Vert \boldsymbol{\epsilon} - \hat{\boldsymbol{\epsilon}} \Vert^2} {\Vert \boldsymbol{\epsilon} \Vert^2} \right\}$ for CFO estimation, where $\boldsymbol{\epsilon}=[\epsilon_1, \epsilon_2, \ldots, \epsilon_K]^{\rm{T}}$ and $\boldsymbol{\hat{\epsilon}}=[\hat{\epsilon}_1, \hat{\epsilon}_2, \ldots, \hat{\epsilon}_K]^{\rm{T}}$, and $\eta_{\rm{g}} =\frac{1}{MK} \mathbb{E}\left\{ \frac{\Vert \boldsymbol{G}_{k,m} \widehat{\boldsymbol{G}}_{k,m} \Vert^2} {\Vert \boldsymbol{G}_{k,m} \Vert^2} \right\}$ for CIR estimation.

Since the estimation method in \cite{RIS_pattern3} was not originally designed with CFO estimation capabilities, we extend it by adding the following CFO estimation approach. Only one user sends its periodic pilot sequence of period $L$ for CFO estimation. While the pilot sequences for CFO estimation are being transmitted, the RIS reflection coefficients are constant. By correlating the samples between the received pilot sequences assigned for each user, the CFO is estimated as in (\ref{eq::c_(kappa)}). For a fair comparison, the same number of pilot resources, i.e., $2L(R+1)$, are used for the proposed method and for the method of \cite{RIS_pattern3}. The length of the extra overhead is approximately proportional to $R$. Since a large value of $R$ is used in RIS systems in practice, the additional resources required for CFO estimation can result in the estimates being outdated. Also, the pilot resources required for \cite{RIS_pattern4} are double those required for the proposed joint CFO and CIR estimation method. Consequently, we can say the proposed method shows the best pilot resource efficiency.


\begin{figure}[t]
    \centering
    \includegraphics[scale=0.5]{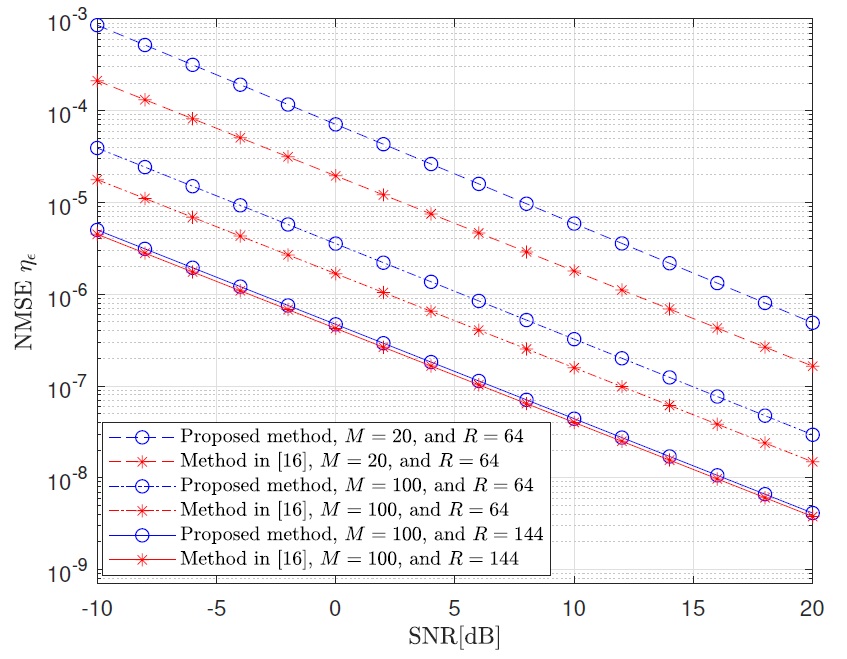}
    \caption{NMSE performance of the proposed CFO estimation method and of the approach in \cite{RIS_pattern4} as a function of SNR, for different values of $M$ and $R$.}
    \label{P1_CFO_MSE}
\end{figure}

In Fig. \ref{P1_CFO_MSE}, it can be seen that the NMSE performance of the proposed CFO estimation method and of the approach in \cite{RIS_pattern4} both improve with the number of BS antennas $M$ and RIS elements $R$. Also, it can be observed that the performance gap between the CFO estimation method of the method in \cite{RIS_pattern4} and the proposed CFO estimation method is reduced when $M$ and $R$ become larger. This means that as $M$ and $R$ increase, the NMSE is dominated by the interference and the effect of noise is by comparison negligible. If $M$ and $R$ are sufficiently large, the proposed method has a performance that is very close to that of \cite{RIS_pattern4}, while using only half of the pilot resources.


\begin{figure}[t]
    \centering
    \includegraphics[scale=0.5]{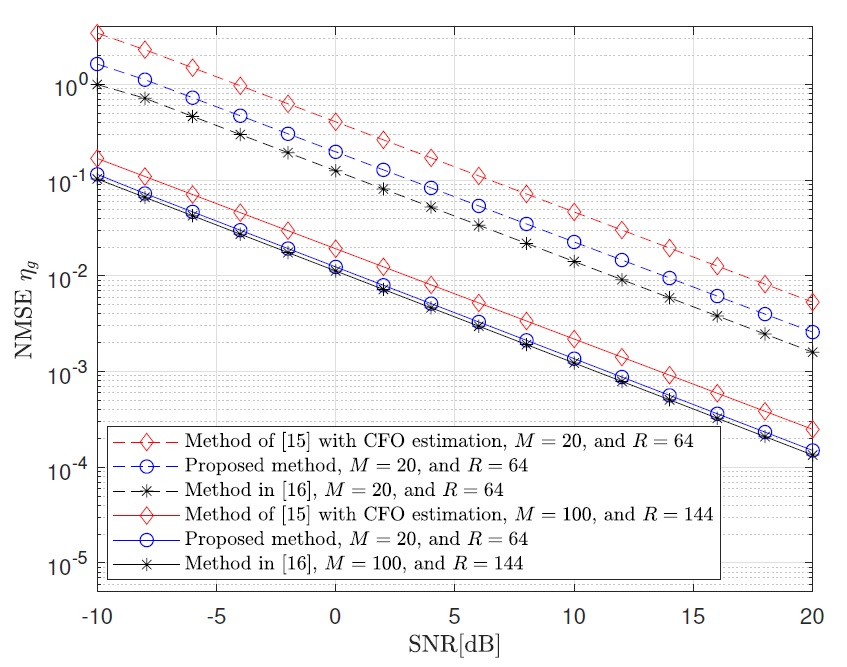}
    \caption{NMSE performance of the proposed CIR estimation method and of the methods in \cite{RIS_pattern3} and \cite{RIS_pattern4} as a function of SNR, for different values of $M$ and $R$.}
    \label{P2_CIR_MSE}
\end{figure}

In Fig. \ref{P2_CIR_MSE}, we show the NMSE performance of different CIR estimation methods for different values of $M$ and $R$. As shown in Fig. \ref{NMSE_CIR_as_CFO}, the NMSE performance of CIR estimation improves as the CFO variance decreases. Consequently, the NMSE performance of CIR estimation also improves. The NMSE performance of CIR estimation for the proposed method shows worse performance compared to the method in \cite{RIS_pattern4}; however, it can be observed the CIR estimation performance of the proposed method becomes close to that of the method in \cite{RIS_pattern4} as $M$ and $R$ increase. This is because the CFO estimation accuracy improves as $M$ and $R$ increase. Consequently, there is a tradeoff between the NMSE performance and the pilot resource efficiency. Also, because of the extra CFO estimation step, the method of \cite{RIS_pattern3} requires more pilot resources than our proposed method. Nevertheless, the proposed method provides an improved NMSE performance compared to the method of \cite{RIS_pattern3}. This is because the MUI caused by residual CFOs degrades the performance of the channel estimation method in \cite{RIS_pattern3}.


\begin{figure}[t]
    \centering
    \includegraphics[scale=0.5]{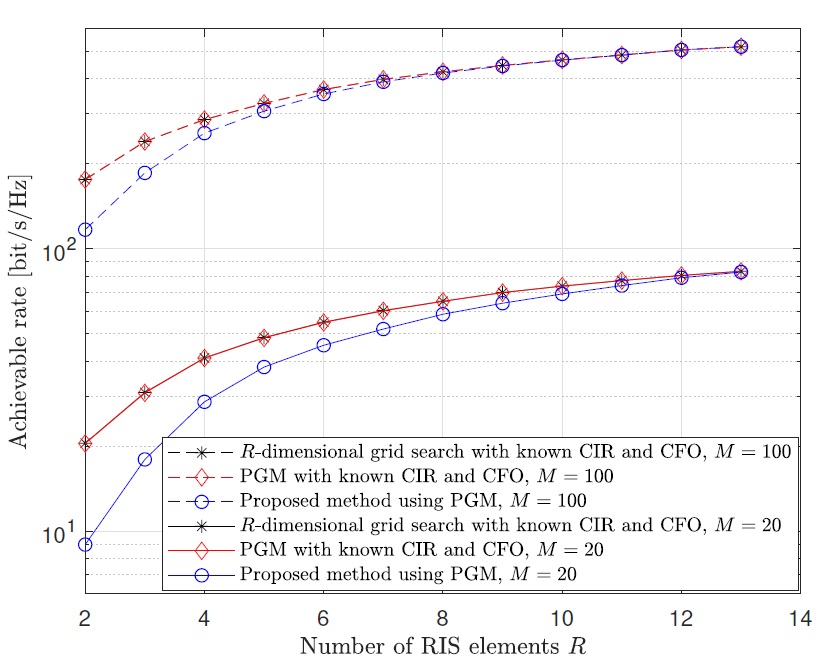}
    \caption{Achievable rate performance of the proposed PGM as a function of $R$ with different values of $M$ when SNR$=10$ dB.}
    \label{AR_R}
\end{figure}

\begin{figure}[t]
    \centering
    \includegraphics[scale=0.5]{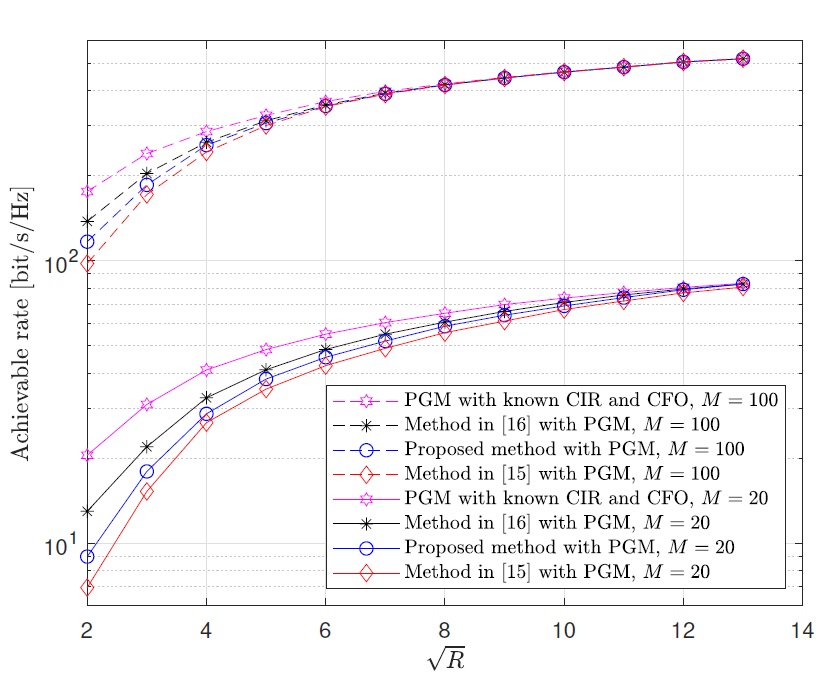}
    \caption{Achievable rate performance of the methods in \cite{RIS_pattern3,RIS_pattern4} and of the proposed joint CFO/CIR estimation method as a function of $R$ with different values of $M$, when SNR$=10$ dB.}
    \label{AR_R_Comp}
\end{figure}

Fig. \ref{AR_R} shows the achievable rate performance of the proposed PGM when SNR$=10$ dB. As benchmarks for the proposed PGM, we also plot the optimal achievable rate obtained by using an $R$-dimensional grid search to maximize (P1) and the achievable rate obtained by the proposed PGM when the CFOs and CIR matrices are assumed to be perfectly known. As shown in Fig. \ref{AR_R}, when the CFOs and CIR matrices are accurately estimated, the achievable rate obtained by the proposed PGM is indistinguishable from the optimal achievable rate obtained by the $R$-dimensional grid search. Also, the achievable rate performance of the proposed PGM with the proposed joint CFO and CIR estimation method becomes close to the optimal achievable rate as $M$ and $R$ increase. This is because when $M$ and $R$ increase, the accuracy of the proposed CFO estimation method increases, and consequently, that of the proposed CIR estimation method also increases. Based on this result, the proposed joint CFO and CIR estimation method with the proposed PGD can achieve a close to optimal achievable rate when $M$ and $R$ have large values which is the case in RIS-aided massive MIMO systems. Also, Fig. \ref{AR_R_Comp} shows the achievable rate performance comparison of the proposed joint CFO and CIR estimation method to that of the methods in \cite{RIS_pattern3,RIS_pattern4}. Based on the result in Fig. \ref{P2_CIR_MSE}, it is observed that as the NMSE performance of CIR estimation improves (i.e., when $R$ increases), the achievable rate performance is close to the optimal achievable rate performance, which is obtained by the grid search.


\begin{figure}[t]
    \centering
    \includegraphics[scale=0.5]{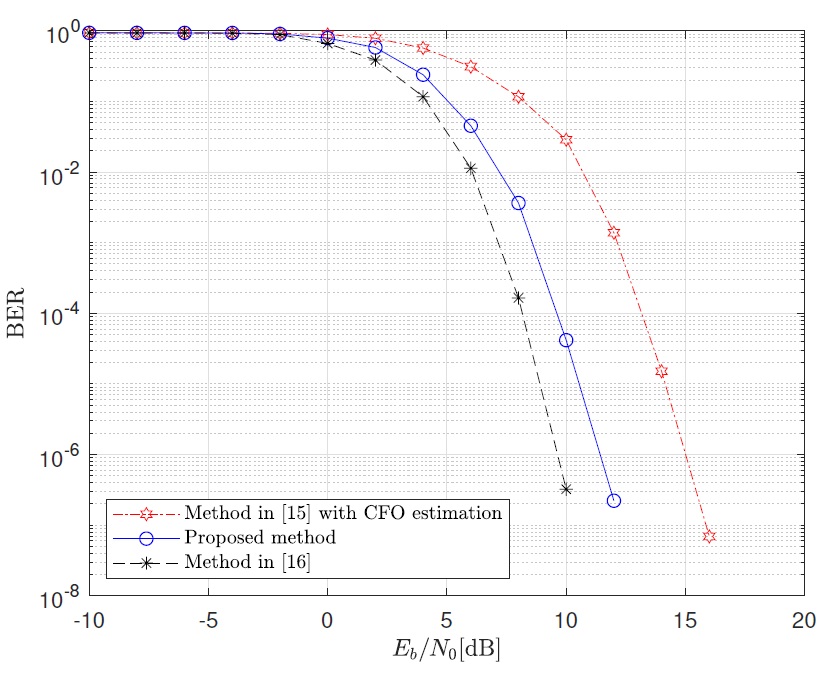}
    \caption{BER performance of the proposed joint CFO and CIR estimation method compared to that of the approaches of \cite{RIS_pattern3} and \cite{RIS_pattern4} as a function of $E_{b}/N_{0}$.}
    \label{BER}
\end{figure}

To evaluate the overall performance of the system using both the proposed CFO and CIR estimation method and the proposed PGM, in Fig. \ref{BER} we show the bit error rate (BER) performance of the methods in \cite{RIS_pattern3} and \cite{RIS_pattern4}, and of the proposed method as a function of $E_{b}/N_{0}$, which denotes the ratio of energy per bit to noise power spectral density. In this simulation, SNR$=10$ dB, $K=3$, $M=20$, $R=64$, and $L=32$. In the CFO and CIR estimation phase, the methods in \cite{RIS_pattern3,RIS_pattern4} and the proposed method use $L(K+2)(R+1)$, $2KL(R+1)$, and $KL(R+1)$ pilot resources, respectively. Using the proposed PGM with $\boldsymbol{\widehat{G}}_{k,m}$ obtained by each estimation method, the RIS reflection coefficient vector $\boldsymbol{\phi}_{\rm{d}}$ is optimized. In the data transmission phase, the $K$ users send 16-ary phase-shift keying (PSK) modulated OFDMA symbols of length $N/K$ for $R+1$ blocks. Here an interleaved subcarrier allocation for each user is utilized. As shown in Fig. \ref{BER}, while the method in \cite{RIS_pattern4} shows the best BER performance among the three methods, the proposed method shows a better performance than the method of \cite{RIS_pattern3}. This trend of the BER performance in Fig. \ref{BER} is similar to that of the NMSE performance in Fig. \ref{P2_CIR_MSE}. The reason is that the error in channel equalization increases as the CIR estimation performance becomes worse. It is worth noting that the proposed method utilizes half the amount of pilot resources than that for the method in \cite{RIS_pattern4} and also fewer pilot resources than that of the method of \cite{RIS_pattern3}. 

\begin{figure}[t]
    \centering
    \includegraphics[scale=0.5]{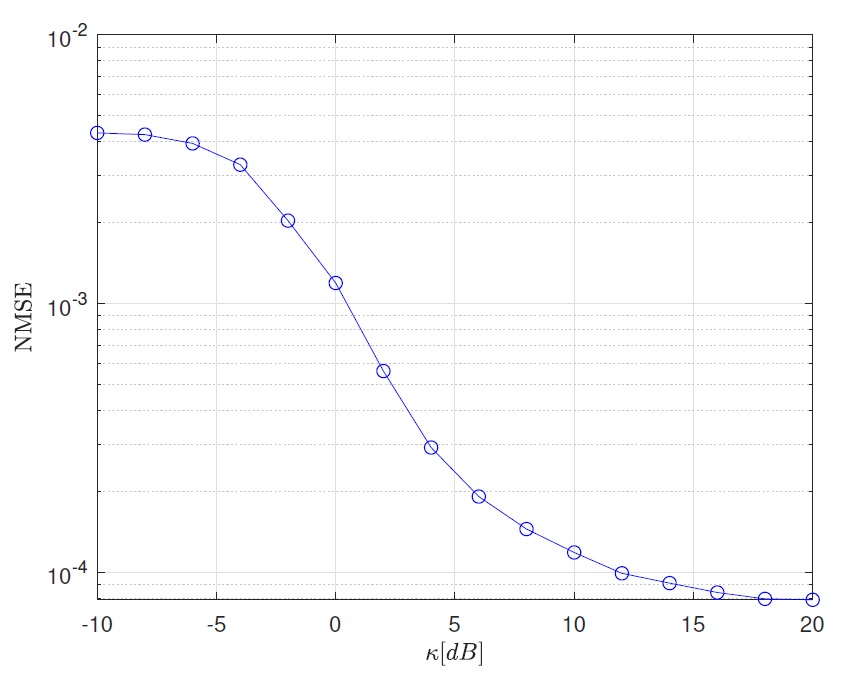}
    \caption{NMSE performance of the proposed joint CFO and CIR estimation method as a function of $\kappa$. Here SNR$=10$ dB, $k=3$, $L=32$, $M=100$ and $R=64$.}
    \label{NMSE_kappa}
\end{figure}

Fig. \ref{NMSE_kappa} shows the NMSE performance for the proposed joint CFO and CIR estimation method as a function of $\kappa$. As $\kappa$ increases, there are three regions: i) a first error floor in the range of $[-10, -4]$ dB; decreasing region in the range of $[-6, 12]$ dB; and a second error floor in the range of $[14, 20]$ dB. With this result, it is observed that the multi-path interference degrades the NMSE performance of channel estimation. Especially in regions (i) and (ii), the NMSE performance of the proposed CIR estimation method is more influenced by the interference from non-deterministic channel components than by the noise. 

\section{Conclusion}

In this paper, we have demonstrated the deleterious effect of CFO on the NMSE performance of LS channel estimation for OFDM-based RIS-aided multi-user massive MIMO systems. We have proposed, for the first time in the literature, a joint CFO and CIR estimation method for such systems. The proposed pilot structure allows for the estimation of the CFO for each user without MUI. With the obtained CFO estimates, the CIR matrix is then estimated using LS estimation. We demonstrated that the proposed joint estimation method exhibits a similar performance in the NMSE and the BER to that of a TDMA-based approach, but requires only half of the pilot overhead. Moreover, the proposed estimation method shows clearly better performance compared to a method using OFDMA while simultaneously requiring a lower overhead. Finally, we proposed a low-complexity PGM for the RIS reflection optimization which provides approximately the same performance as the more computationally demanding grid search method.


\end{document}